\documentclass[pre,twocolumn,superscriptaddress,nobibnotes]{revtex4-2}
\usepackage{amsmath,amssymb,amsfonts,latexsym}\usepackage{amsmath,bm}
\usepackage{graphicx}
\usepackage{mathptmx}
\usepackage[nomessages]{fp}
\usepackage{mathrsfs} 

\DeclareMathAlphabet{\mathcal}{OMS}{cmsy}{m}{n}
\usepackage{xcolor}
\usepackage{bbding}
\newcommand{\e}{{\rm e}}
\usepackage{braket}
\usepackage[protrusion=true,expansion=true]{microtype}
\usepackage{times}
\usepackage{hyperref}\hypersetup{
    pdftitle={main},    
    pdfauthor={},
    colorlinks=true,
    citecolor=blue,
    linkcolor=blue,
    urlcolor=blue
  }

\newcommand{\E}{{\mathbb{E}}}

\newcommand{\de}{\partial}
\newcommand{\be}{\begin{equation}}
\newcommand{\ee}{\end{equation}}
\newcommand{\nn}{\nonumber\\[4pt]}

\newcommand\tr{\text{tr}}
\newcommand{\n}{\mathfrak{n}}
\newcommand{\jj}{\mathfrak{j}}
\newcommand\g{\mathfrak{g}}

\newcommand\cum{\mathfrak{c}}
\newcommand\p{\mathfrak{p}}

\newcommand{\sign}{{\varepsilon}}

\definecolor{red}{rgb}{0.8,0,0.15}
\definecolor{blue}{rgb}{0.15, 0.15, .8}
\definecolor{green}{rgb}{0., 0.5, .4}
\newcommand{\titleinfo}{Universal classical and quantum fluctuations \\ in the large deviations of current of noisy quantum systems:\\ Quantum symmetric simple exclusion and
inclusion processes}
\begin{document}
\title{\titleinfo}
\author{Mathias Albert}
\affiliation{Universit\'e Côte d’Azur, CNRS, Institut de Physique de Nice, 06200 Nice, France}
\affiliation{Institut Universitaire de France (IUF)}
\author{Denis Bernard}
\affiliation{Laboratoire de Physique de l’\'Ecole Normale Superieure, CNRS, ENS \& Universit\'e PSL,
Sorbonne Universit\'e, Universit\'e Paris Cit\'e, 75005 Paris, France}
\author{Tony Jin}
\affiliation{Universit\'e Côte d’Azur, CNRS, Centrale Med, Institut de Physique de Nice, 06200 Nice, France}
\author{Stefano Scopa$^3$}
\email{stefano.scopa@phys.ens.fr}
\author{Shiyi Wei$^1$}
\begin{abstract}
We study the fluctuation statistics of integrated currents in noisy quantum diffusive systems, focusing on the Quantum Symmetric Simple Exclusion and Inclusion Processes (QSSEP/QSSIP). These one-dimensional fermionic (QSSEP) and bosonic (QSSIP) models feature stochastic nearest-neighbor hopping driven by Brownian noise, together with boundary injection and removal processes. They provide solvable microscopic settings in which quantum coherence coexists with diffusion. Upon noise averaging, their dynamics reduce to those of the classical SSEP/SSIP. We show that the cumulant-generating function of the integrated current, at large scales, obeys a large deviation principle. To leading order in system size and for each noise realization, it converges to that of the corresponding classical process, establishing a classical typicality of current fluctuations in these noisy quantum systems. We further demonstrate a direct connection with Macroscopic Fluctuation Theory (MFT), showing that the large-scale equations satisfied by biased quantum densities coincide with the steady-state Hamilton equations of MFT, thereby providing a microscopic quantum justification of the MFT framework in these models. Finally, we identify the leading finite-size corrections to the current statistics. We show the existence of subleading contributions of purely quantum origin, which are absent in the corresponding classical setting, and provide their explicit expressions for the second and third current cumulants. These quantum corrections are amenable to direct experimental or numerical verification, provided sufficient control over the noise realizations can be achieved. Their presence points toward the necessity of a quantum extension of MFT.
\end{abstract}

\maketitle
\date{\today}
\section{Introduction}
Exploring the emergence of universal behavior in the fluctuation statistics of nonequilibrium quantum systems has become a central topic in contemporary statistical physics. While the dynamics of isolated or weakly interacting systems is often governed by unitary evolution, realistic quantum systems are invariably affected by noise and interactions with their environment. {In the classical realm}, stochastic models have long provided paradigmatic examples for understanding transport and fluctuations. In particular, the Symmetric Simple Exclusion Process (SSEP) and related models offer a well-defined framework to compute exact large deviation functions, characterize rare events, and formulate Macroscopic Fluctuation Theory (MFT) for diffusive transport~\cite{Kipnis1989,Spohn1991,derrida1993exact,Derrida2007,bodineau2004current,Mallick2015,bertini2001fluctuations,bertini2002macroscopic}. 
These models have proven invaluable for identifying universal features in the statistical properties of transport in nonequilibrium systems.\\
\indent
However, in the quantum realm, the interplay between coherence, environmental interactions, and noise makes it crucial to understand whether classical descriptions remain valid and how quantum fluctuations manifest at mesoscopic scales~\cite{Bernard2021}. Tractable models such as the Quantum Symmetric Simple Exclusion Process (QSSEP)~\cite{Bauer2019,Bernard2019} and its recent extensions~\cite{Bernard2025,alba2025nuqssep,russotto2025,Bernard2026-domain,iqsep-paper} have enabled detailed studies of fluctuations in noisy quantum systems. Remarkably, in these models, averaging over noise realizations reproduces classical stochastic dynamics—for instance, the noise-averaged dynamics of QSSEP reduces to SSEP—while individual realizations retain coherent quantum features. These quantum formulations reveal elegant connections \cite{Hruza2023,Bauer2024} with free probability theory~\cite{V97,Mi17,S19} and offer alternative routes to exact large deviation calculations even beyond classically integrable settings. A notable example is the solution of the Quantum Symmetric Inclusion Process (QSSIP)~\cite{Bernard2025}, whose noise-averaged dynamics is governed by the classical SSIP~\cite{Giardina2010,Grosskinsky2011} and does not map onto integrable spin chains via standard methods.\\
\indent
These findings point toward a striking form of universality: for large systems and long times, the full counting statistics and large deviation functions of transport in noisy quantum systems are effectively classical, despite the persistence of quantum coherences at the microscopic level~\cite{Bernard2021,Hruza2023}. More specifically, it has been conjectured that in noisy quantum diffusive systems, transport fluctuations satisfy a large deviation principle realization-wise, and that the corresponding large deviation functions are, to leading order in system size, independent of the noise realization and coincide with those of the associated classical process defined by the mean dynamics. Evidence supporting this conjecture was obtained early on --although not fully appreciated at the time-- in the context of diffusive charge transport in quasi-1d disordered materials \cite{LeeLevitovYuPRBUniversalstatistics,Derrida2004,HruzaJinAndersonQSSEP} and has also recently been obtained in charge-conserving random circuits~\cite{FCSMFTDeNardis}. It should be noted that this conjecture pertains to the leading contribution to transport fluctuations and does not extend to their subleading correction and to fluctuations of the quantum coherences, which remain intrinsically non-classical. In Refs.~\cite{Bauer2024,Bernard2025}, the classical typicality of large deviations for density fluctuations was investigated for QSSEP and QSSIP. In this work, we complement these studies by analyzing the integrated current statistics and the corresponding large deviation function of QSSEP/QSSIP models.\\

In formulating this problem, one is immediately confronted with the well-known difficulty of defining and measuring currents in quantum systems. First, integrated currents are not associated with local observables which can be projectively measured. Second, any measurement protocol necessarily induces a back action on the system, which must be consistently incorporated in the description. The strategy adopted here is to extract the transferred charge across a given link, $Q_j(t)$, from two measurements of the particle density in auxiliary baths at times $0$ and $t>0$ (see Fig.~\ref{fig:setup}). Summing over the outcomes of the two measurements, one obtains a well-defined moment-generating function
\[
Z_{t,w}(u)=\sum_{\{Q_j(t)\}} e^{-u\sum_j \lambda_j Q_j(t)} \,\mathrm{Prob}(\{Q_j(t)\}),
\]
from which, by virtue of the continuity equation, the statistics of the integrated current can be inferred. This provides a consistent protocol to access current fluctuations, known as the two-time measurement scheme, which has been widely used in the literature, see e.g.~Refs.~\cite{two_times_kurchan2001quantumfluctuationtheorem,two_times_tasaki2000jarzynskirelationsquantumsystems,two_times_Esposito_2009,Landi2024current}. Biasing the bulk dynamics with amplitude $u$ is accompanied by the introduction of auxiliary degrees of freedom $\lambda_j$ associated with each link of the chain. As first noticed in Ref.~\cite{Derrida2004,Derrida2007} and more recently emphasized in Ref.~\cite{costa2025}, this gauge freedom allows for non-perturbative treatments of the problem. Alternative approaches to current statistics consist in biasing (or ``tilting'') the quantum jump operators in the boundary-driven Lindblad dynamics, e.g.~\cite{Garrahan2010,nidari2014,nidari2014b,Carollo2017,Monthus2017,Carollo2018}. In Sec.~\ref{sec:2time-meas}, we show that this widely used prescription is in fact equivalent to the present construction.

It is important to stress that the noisy dynamics of the quantum chain renders $Z_{t,w}(u)$ itself a fluctuating quantity, which depends on the specific realization of the noise, denoted throughout by the subscript `$w$'. By construction, after averaging over the noise, $\mathbb{E}[Z_{t,w}(u)]$ generates the moments of the associated classical SSEP/SSIP stochastic process. Using exact calculations based on the properties of the two-point matrices $G^{(w)}$ governing the nonequilibrium steady state of the biased system (introduced in Refs.~\cite{Bauer2019, Bernard2019} in absence of bias), we further show that $Z_{t,w}(u)$ obeys a large deviation principle,
\[
Z_{t,w}(u)\underset{N\to\infty}{\asymp} e^{N \tau F_{{\rm qu},w}(u)},
\]
for times scaling diffusively as $t=\tau N^2\sim {\cal O}(N^2)$, and that the corresponding large deviation function is self-averaging: for a typical realization of the noise, it converges to its noise-averaged (classical) value, with corrections that are subleading in system size. More precisely, we show that
\[
\lim_{N\to\infty}\frac{\mathbb{E}[\log Z_{t,w}(u)]}{N\tau}
\sim
\lim_{N\to\infty}\frac{\log \mathbb{E}[Z_{t,w}(u)]}{N\tau},
\]
where the equivalence between the two cumulant-generating functions holds asymptotically at large system size $N$. For QSSEP, this result implies that the current statistics, to leading order in system size, reproduces that of the SSEP~\cite{Derrida2004,bodineau2004current,Derrida2007}, as recently observed in Ref.~\cite{costa2025}. For QSSIP, to the best of our knowledge, the large deviation function of the current in the underlying classical process has not previously been obtained analytically, although numerical results for asymmetric inclusion processes have been reported in Ref.~\cite{Minoguchi2023}.\\

Our derivation of the cumulant-generating function in noisy quantum models also reveals a striking connection with MFT. We show that the large-scale properties of the generating function can be determined solely from the biased mean density $\mathbb{E}[G^{(w)}_{ii}]$, for a suitable choice of the weights $\lambda_i$ associated with each link~\cite{costa2025}. Remarkably, the equations satisfied in the large-$N$ limit by the scaling functions $\g_1(x=i/N)\sim \mathbb{E}[G^{(w)}_{ii}]$ and $\xi(x=i/N)\sim N\lambda_i$ are analogous to the Hamilton equations arising in MFT. Within this correspondence, the biased density $\g_1(x)$ of the quantum model plays the role of the optimized density profile in MFT, while the gauge field $\xi(x)$ is related to the response field in the path-integral formulation.\\

This correspondence provides a microscopic explanation for the observed classical typicality of transport fluctuations at large scales and offers, in turn, a microscopic justification of MFT from the quantum dynamics embedding the classical diffusive model. Building on this result, we propose a variational ansatz for the quantum problem. The central idea is to exploit the classical reduction to SSEP/SSIP after noise averaging and to infer the cumulant-generating function from the biased Markov generator, as already done in Refs.~\cite{Donsker1975,Lebowitz1999,Jack2010,Derrida2007,Chetrite2014}. We show, however, that the same strategy can be applied directly to the full quantum dynamics, reducing the problem to the determination of a growth-rate exponent, akin to a Lyapunov exponent, of a linear stochastic problem. This yields
\[
\frac{\log Z_{t,w}(u)}{t} \underset{t\to\infty}{\asymp}\underset{{\hat L,\hat R}}{\text{extr}}\, \langle \hat L | \mathcal{L}_{-u} | \hat R \rangle,
\]
where {the extremum sets $\hat L$ and $\hat R$ equal to the principal} left and right eigenstates of the (non-Hermitian) biased Lindbladian $\mathcal{L}_{-u}$. We show that the outcome of this variational principle coincides with the results obtained from exact calculations and from MFT, while holding for each {typical} realization of the noise. In this sense, the classical typicality of current fluctuations is not merely an averaged property: there is an asymptotic convergence, realization by realization, toward the classical cumulant-generating function.\\

Having established the classical character of current statistics at leading order, a natural question is whether genuinely quantum effects survive in the subleading corrections. Of course, quantum effects are present in the coherence statistics, so the question here is whether quantumness survives also in transport statistics. In the scaling limit $N\to\infty$ with $t\sim N^2$, we show the presence of corrections to the current statistics of the noisy quantum chain at order ${\cal O}(N^{-1})$, of purely quantum origin. We relate this leading quantum contribution to the scaled coherence-loop function $\g_2(x=i/N,y=j/N)\sim N\,\mathbb{E}[G^{(w)}_{ij}G^{(w)}_{ji}]$. We then derive the equation satisfied by $\g_2$ and explicitly compute the resulting quantum corrections to the variance and skewness of the current cumulant distribution. More generally, we discuss the structure and closure of the hierarchy of equations satisfied by the scaled loop functions $\g_n$ at arbitrary order, which in principle allows for the exact computation of quantum corrections to current statistics to any order. {Although these corrections enter only as subleading contributions to transport fluctuations, their importance becomes apparent in concrete observables, for instance the current variance. In particular, the simplest experimental or numerical protocol to access the current variance is to consider multiple samples and average over both types of averages, i.e., $\big(\E[\langle Q^2 \rangle]-\E[\langle Q\rangle]^2\big)/t$, where $\langle \ \rangle$ denotes the quantum expectation introduced below. This protocol reproduces the known classical results, for example those found in SSEP~\cite{Derrida2004, Bodineau2007, AppertRolland2008}. If, however, experimental control over the noise realizations is available, one could instead measure the current variance for a fixed noise realization and then take its statistical average over realizations, namely $\E[\langle Q^2(t) \rangle-\langle Q(t)\rangle^2]/t$. Beyond leading order, the two protocols differ, thereby revealing the presence of measurable quantum corrections to the current statistics, which we quantify in this work. Moreover, their existence highlights the limitations of classical MFT for a complete description of diffusive quantum systems at mesoscopic scales. This observation points toward the necessity of formulating a quantum extension of MFT \cite{Bernard2021}.}\\

{Finally, we note that similar quantum corrections appear in mesoscopic physics, underlying the phenomenon of ``{\it Universal Conductance Fluctuations}'' (UCF)~\cite{StoneUCF1985,WebbUCF1985,MaillyUCF1992,LeeLevitovYuPRBUniversalstatistics,Beenakker1997}, which encode information about symmetry classes beyond average transport properties. Ref.~\cite{Derrida2004} showed that the leading-order current cumulants of the SSEP exactly coincide with those found in quasi-1D free fermionic systems in the metallic regime with quenched disorder~\cite{LeeLevitovYuPRBUniversalstatistics}, highlighting a remarkable correspondence between classical stochastic and mesoscopic quantum transport. Whether there exists a relation between the UCF and the QSSEP/QSSIP quantum corrections remains an open question that we leave for future work.}

\vspace{0.2cm}
{\it Outline}.~---~The paper is organized as follows. In Sec.~\ref{sec:model}, we recall the definition and basic properties of the QSSEP/QSSIP models. In Sec.~\ref{sec:2time-meas}, we outline the two-time measurement protocol used to compute the current statistics and present the corresponding biased evolution for the system density matrix. Section~\ref{sec:LDF} introduces the large deviation function of currents for both the quantum and classical (noise-averaged) processes, highlighting the main results on classical typicality and quantum corrections discussed in the following sections. Section~\ref{sec:exact-calc} presents the general strategy and the assumptions used for the calculation of the current statistics. In Sec.~\ref{sec:classical}, we focus on the leading contribution to the cumulant-generating function, showing that it coincides with that of the underlying classical process, and commenting on the physical implications for the bosonic case. We also explicitly demonstrate, in Sec.~\ref{sec:classical-MFT}, that the biased quantum dynamics reproduces, to leading order in the system size, the same equations obtained within MFT. In Sec.~\ref{sec:var}, {we establish the self-averaging of the current cumulant-generating function for a typical realization of the noise, and formulate the corresponding variational problem.} Sec.~\ref{sec:quant-fluc} is devoted to the lowest-order calculation of the quantum correction to the current cumulants, and Sec.~\ref{sec:conclusion} provides a brief summary and conclusion. A collection of appendixes~\ref{app:two-time-measurem}-\ref{app:details-extr-variational-app} contains additional details and technical derivations of the results presented in the main text.
\section{The model}\label{sec:model}
We consider a one-dimensional quantum chain with nearest neighbor stochastic hoppings, whose bulk dynamics is generated by the stochastic Hamiltonian increment
\be\label{eq:dH}
d\hat{H}_w=\sum_{j=0}^{N-1} \left(dw_j(t) \hat\ell_j + d\overline{w}_j(t) \hat\ell_j^\dagger\right),
\ee
where $\hat\ell_j:= \hat{c}^\dagger_{j+1}\hat{c}_j$ and $\hat{c}^\dagger_j$, $\hat{c}_j$ are the creation and annihilation operators of fermionic or bosonic particles, satisfying the following commutation relations
\be
[\hat{c}_i;\hat{c}_j^\dagger]_\sign := \hat{c}_i\hat{c}_j^\dagger -\sign \hat{c}^\dagger_j \hat{c}_i=\delta_{ij}
\ee
and otherwise zero. Here, $\sign=\pm 1$ for bosons or fermions, respectively, and $w_j(t)$, $\overline{w}_j(t)$ are pairs of complex conjugated Brownian motions, with nonvanishing quadratic variations $dw_j(t) d\overline{w}_k(t)= dt \delta_{jk}$.\\

 At the endpoints of the chain, we couple the system to dissipative reservoirs. Assuming a Markovian interaction with the leads, {the stochastic evolution of the density matrix reads}
\be\label{eq:quant-dyn} 
d\hat\rho_{t,w}=-i[d\hat{H}_w;\hat\rho_{t,w}] + dt {\cal L}(\hat\rho_{t,w}) + dt {\cal L}_\text{bdry}(\hat\rho_{t,w}),
\ee
where $dt\,{\cal L}(\bullet):=-\frac12[d\hat H_w;[d\hat H_w;\bullet]]$ {is the deterministic It\=o contribution arising at order ${\cal O}(dt)$ from the stochastic unitary evolution $e^{-id\hat H_w}\hat\rho_{t,w}e^{id\hat H_w}$, while
${\cal L}_\text{bdry}$ describes the Lindblad coupling to the reservoirs,}
\begin{align}\label{eq:Lbdry}
{\cal L}_\text{bdry}(\bullet)=\sum_{p\in\{0,N\}}\!\!&\Big[\alpha_p \big(\hat{c}^\dagger_p \bullet \hat{c}_p-\frac12\{ \hat{c}_p\hat{c}_p^\dagger;\bullet\}\big )\nn
&+\beta_p\big(\hat{c}_p \bullet \hat{c}^\dagger_p-\frac12\{ \hat{c}^\dagger_p\hat{c}_p;\bullet\}\big )\Big].
\end{align}
$\alpha_p$ and $\beta_p$ are model parameters controlling the injection and extraction of particles at sites $0$ and $N$. In the limit $N \to \infty$, these parameters fix the boundary densities of the chain, without fluctuations, as
\be\label{eq:baths-dens-values}
\bar{n}_0 = \frac{\alpha_0}{\beta_0 - \sign \alpha_0}, \quad
\bar{n}_N = \frac{\alpha_N}{\beta_N - \sign \alpha_N},
\ee
with $\beta_p > \alpha_p$ required in the bosonic case ($\sign = +1$) to reach a steady state.  Depending on the quantum statistics of the particles (bosonic or fermionic, corresponding to $\sign = \pm 1$), 
this setup is referred to as the `{open QSSIP}' \cite{Bernard2025} or `{open QSSEP}' \cite{Bernard2019}, respectively.\\

It is also well known that, considering the noise-averaged process $\bar\rho_{t} := \E[\hat\rho_{t,w}]$, where $\E[\bullet ]$ denotes the statistical average over stochastic realizations (labeled by the subscript $w$ throughout the text), the dynamics \eqref{eq:quant-dyn} reduces to~\cite{Bernard2018}
\be\label{eq:mean-dyn}
\frac{d}{dt}\bar\rho_t= {\cal L}(\bar\rho_t)+{\cal L}_\text{bdry}(\bar\rho_t).
\ee
Since averaging over noise realizations results in a loss of information, the average dynamics is dissipative rather than unitary. Decoherence then causes the off-diagonal elements of $\bar\rho_t$ to be exponentially suppressed in time. The resulting diagonal density matrix, at large times $\bar\rho_{t\gg 1}$, specifies a probability distribution of classical configurations $\bm{n}=(n_0,\dots,n_N)$, namely
\be
\bar\rho_{t}\overset{t\gg 1}{\simeq} \sum_{\bm{n}} P_t(\bm{n}) |\bm{n}\rangle\!\langle \bm{n}|, \quad \sum_{\bm{n}}P_t(\bm{n})=1, \; P_t(\bm{n})\geq 0.
\ee
The bulk dynamics induced by the Lindbladian in \eqref{eq:mean-dyn} acts as
\begin{align}
{\cal L}(|\bm{n}\rangle\!\langle \bm{n}|)=\sum_{j=0}^{N-1} &\gamma_{j\to j+1}\Big(\underset{j\to j+1}{|\bm{n}\rangle\!\langle \bm{n}|} -|\bm{n}\rangle\!\langle \bm{n}|\Big) \nn
&+ \gamma_{j+1\to j}\Big(\underset{j+1\to j}{|\bm{n}\rangle\!\langle \bm{n}|} -|\bm{n}\rangle\!\langle \bm{n}|\Big).
\end{align}
Here $\underset{j\to k}{|\bm{n}\rangle\!\langle \bm{n}|}$ denotes the classical state obtained from ${|\bm{n}\rangle\!\langle \bm{n}|}$ after a single particle hops from site $j$ to $k$, and the left/right transition rates are $\gamma_{j\to k}:= n_j(1+\sign n_{k})$. Similarly, at the boundary sites ($p=0,N$) one has
\begin{align}
&{\cal L}_\text{bdry}(|n_p\rangle\!\langle n_p|)=\alpha_p(1+\sign n_p) |n_p+1\rangle\!\langle n_p+1| \nn
&\;+ \beta_p n_p |n_p-1\rangle\!\langle n_p-1| -\left[\alpha_p+(\beta_p+\sign \alpha_p)n_p \right]|n_p\rangle\!\langle n_p|.
\end{align}

Thus, the noise-averaged Lindblad dynamics in \eqref{eq:mean-dyn} defines a Markov chain on the classical configurations, equivalent to the classical `SSIP’~\cite{Giardina2010,Grosskinsky2011} or `SSEP’~\cite{Derrida2007} process for $\sign=\pm 1$, respectively.

\section{Current statistics from two-time measurement scheme}\label{sec:2time-meas}
For $\bar{n}_{0}>\bar{n}_{N}$ in \eqref{eq:baths-dens-values}, a net particle current flows from left to right (and vice versa when $\bar n_{N}>\bar n_{0}$). As outlined in the introduction, our goal is to characterize this current and its fluctuations. Accessing the integrated current in a single experimental run is generally not feasible, since it does not correspond to a conserved quantity that can be  directly measured.  Moreover, in quantum systems any measurement inevitably induces back action, which requires us to specify clearly the protocol used to extract the current. Both issues have been extensively discussed in the mesoscopic-transport literature, see e.g. \cite{Meso_Levitov_1996,Meso_Nazarov2003,Meso_Clerk,Landi2024current}. A widely used resolution is provided by the two-time measurement scheme~\cite{two_times_kurchan2001quantumfluctuationtheorem,two_times_tasaki2000jarzynskirelationsquantumsystems,two_times_Esposito_2009,Landi2024current}, which we briefly review below.\\
\begin{figure}
    \centering
    \includegraphics[width=.95\columnwidth]{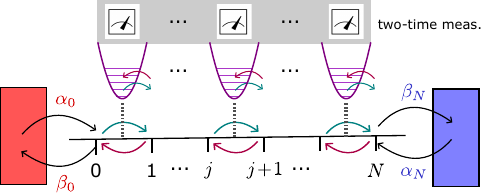}
    \caption{Illustration of the setup. A quantum chain is driven into a nonequilibrium steady state by connecting its endpoints to Markovian reservoirs. Each link is coupled to an auxiliary bath, such that whenever a particle hops from $j\to j+1$ (resp.~$j+1\to j$), a photon is emitted (resp.~absorbed) in that bath. By performing two measurements of the bath occupation at times $t=0$ and $t$, we infer the integrated current $Q_j(t)$ through link $j$. In the Markovian limit, the moment-generating function of the $Q_j(t)$ is evolved using the biased QSSIP/QSSEP Hamiltonian \eqref{eq:dH-bias}.
}
    \label{fig:setup}
\end{figure}

One introduces auxiliary degrees of freedom acting as statistically independent reservoirs attached to each link of the chain. In this setting, each link is coupled to its own reservoir, and the hopping of a particle from $j\to j+1$ (resp.~$j+1\to j$) is accompanied by the emission (resp.~absorption) of a boson in that reservoir; see Fig.~\ref{fig:setup} for an illustration. Up to boundary terms, the Hamiltonian increment describing this system–reservoir coupling takes the form
\be\label{eq: system-bath-evolution}
d\hat{H}_{\rm SR} = \sum_{j=0}^{N-1} \big(\hat{\ell}_{j}\,\hat{A}_j^{\dagger}(t) + \text{h.c.}\big),
\ee
where $\hat{A}_{j}(t)$, $\hat{A}_{j}^{\dagger}(t)$ denote the bosonic annihilation and creation operators of the reservoirs in the interaction picture. The reservoirs themselves are modelled as an infinite collection of free bosonic modes; their explicit Hamiltonian is given in Appendix~\ref{app:two-time-measurem}. \\

\indent
We now consider the two-measurement protocol. At the initial time $t=0$ the total particle number of in $j$th reservoir is measured. The system then evolves according to \eqref{eq: system-bath-evolution}, and the particle number of the same bath is measured again after some time $t>0$. The difference between the two measurements yields a real number, denoted $Q_j(t)$, which defines the total charge that has crossed the $j$th link between times $0$ and $t$. The collection of all $Q_j(t)$ thus provides a set of classical fluctuating variables, one for each link. The corresponding moment-generating function is    
\begin{equation}
Z_{t}\left(\bm{\zeta}\right):=\sum_{\{Q_j(t)\}} \text{Prob}(\{Q_j(t)\}) \ e^{-\sum_{j}\zeta_{j}Q_j(t)}
\end{equation}
where $\bm\zeta=\left\{ \zeta_{j}\right\}_{j=0}^{N-1}$, and $\text{Prob}(\{Q_j(t)\})$ denotes the probability of recording the set of transferred charges in a single realization of the two-measurement scheme. An equivalent representation of $Z_{t}$ is  
\begin{equation}
Z_{t}(\bm{\zeta})=\tr\left(\hat{\rho}^{\rm tot}_{t}(\bm{\zeta})\right)
\label{eq:protogenerating}
\end{equation}
where $\hat{\rho}^{\rm tot}_{t}(\bm\zeta)$ evolves under the (non-Hermitian) biased dynamics
\begin{equation}
\hat{\rho}^{\rm tot}_{t+dt}(\bm{\zeta})
= e^{-id\hat{H}_{SR}(\bm{\zeta})}\;
\hat{\rho}^{\rm tot}_{t}(\bm{\zeta})\;
e^{id\hat{H}_{SR}(-\bm{\zeta})},
\end{equation}
with  
\begin{equation}\label{eq:bias-bath}
d\hat{H}_{SR}(\bm{\zeta}):
=\sum_{j=0}^{N-1}\big(\hat{\ell}_{j}\ \hat{A}_{j}^{\dagger}(t)\  e^{\zeta_{j}/2}+\hat{A}_{j}(t)\ \hat\ell^\dagger_j\  e^{-\zeta_{j}/2}\big).
\end{equation}
Note that, in the scaling limit $N\to \infty$ at fixed $x=j/N$ and $\tau = t/N^2$, the transferred charges across each link are related to one another
\begin{equation}\label{eq:stat-current}
\frac{Q_j(t)-Q_{k}(t)}{t}\leq\frac{\left|j-k\right|}{t}\sim {\cal O}(N^{-1}),
\end{equation}
so any choice of $\{\zeta_j\}$ leads to the same large-scale generating function. It is therefore convenient to introduce the parametrization $\zeta_j := u\,\lambda_j$ with a set of weights $\lambda_j$ satisfying the gauge condition
\be\label{eq:gauge}
\sum_{j=0}^{N-1} \lambda_j = 1.
\ee
This gauge freedom, noticed in \cite{Derrida2004} and more recently in \cite{costa2025}, will be crucial in the following. 
As shown in Appendix~\ref{app:two-time-measurem}, the Markovian limit of the reservoirs in \eqref{eq:bias-bath} can be recast in terms of Brownian hopping increments $dw_{j}(t)$, $d\overline{w}_j(t)$, producing a biased version of the QSSIP/QSSEP dynamics,
\be\label{eq:dH-bias}
d\hat{H}_{w,u}=\sum_{j=0}^{N-1}\big(\hat{\ell}_{j}\,dw_j(t)\, e^{u\lambda_j/2}
+ \hat\ell^\dagger_j\, d\overline{w}_j(t)\, e^{-u\lambda_j/2}\big),
\ee
which reduces to \eqref{eq:dH} when $u=0$, i.e. in the absence of bias. By writing the QSSIP/QSSEP generating function as 
\be
Z_{t,w}(u):=\tr\big(\hat\rho_{t,w}(u)\big),
\ee
the corresponding biased evolution of the system density matrix, to order ${\cal O}(dt)$, is
\be\label{eq:dyn-bias}
d\hat\rho_{t,w}(u) 
   = d{\cal H}_{w,u}[\hat\rho_{t,w}(u)] 
   + dt\,{\cal L}_u[\hat\rho_{t,w}(u)] 
   + dt\,{\cal L}_\text{bdry}[\hat\rho_{t,w}(u)],
\ee
with stochastic contribution
\begin{align}
d{\cal H}_{w,u}(\bullet) := -i \left(d\hat{H}_{w,u}\ \bullet - \bullet \ d\hat{H}_{w,-u}\right),
\end{align}
and deterministic part
\begin{align}
{\cal L}_u(\bullet) := \sum_{j=0}^{N-1}\Big[ 
   e^{u\lambda_j}\hat\ell_j \bullet \hat\ell^\dagger_j
   + e^{-u\lambda_j}\hat\ell^\dagger_j \bullet \hat\ell_j
   - \tfrac{1}{2}\{\hat\ell_j\hat\ell_j^\dagger + \hat\ell^\dagger_j\hat\ell_j; \bullet\}\Big].
\end{align}
One may also define the noise-averaged biased state as $\bar\rho_t(u):=\E[\hat\rho_{t,w}(u)]$, whose dynamics is given by
\be
\frac{d}{dt}\bar\rho_t(u)= \big({\cal L}_u+{\cal L}_\text{bdry}\big)[\bar\rho_t(u)].
\ee
\section{Large deviation behavior}\label{sec:LDF}
Equipped with these definitions, we are now in a position to address the large deviation functions. In the present setting, the large deviation hypothesis states that, at sufficiently large scales $N\to \infty$ with $\tau=t/N^2$ fixed, the following relation is expected to hold
\begin{equation}\label{eq:def-Fw}
Z_{t,w}(u)\underset{N\to\infty}{\asymp} e^{ N \tau F_{{\rm qu},w}(u)},
\end{equation}
where $F_{{\rm qu},w}(u)$ is the cumulant-generating function of the integrated current for the single realization of the noise, and $\asymp$ means that the equality holds after taking the log on both sides and large $N$. The associated cumulant-generating function for the classical process is, by construction,
\begin{equation}
\mathbb{E}\left[Z_{t,w}(u)\right]\underset{N\to\infty}{\asymp} e^{N\tau F_{{\rm cl}}(u)},
\end{equation}
and for the SSEP ($\sign=-1$), the expression for $F_{\rm cl}(u)$ has been derived in Refs.~\cite{Derrida2004,Derrida2007,Derrida2011}. The analogous expression for SSIP is instead not known. A priori, $F_{{\rm qu},w}(u)$ in eq.~\eqref{eq:def-Fw} could be fluctuating due to the stochastic dynamics. We shall argue that in the large-$N$ limit this is not the case: ${F}_{{\rm qu},w}(u)$ self-averages and converges, for each realization of the noise, to its noise-averaged (or classical) value,
\begin{equation}
F_{{\rm qu},w}(u)\sim F_{{\rm cl}}(u)+{\cal O}(N^{-1}).
\end{equation}
We shall refer to this result as {\it classical typicality} of the large deviation function of currents in the discussion below.\\

We will further isolate the leading quantum corrections to classical typicality by inspecting the quantity
\be\label{eq:def-dFqu}
\delta F_{\rm qu}(u) := \E[F_{{\rm qu},w}(u) - F_{\rm cl}].
\ee
Remarkably, a closely related phenomenology also emerges in the, a priori completely different, context of mesoscopic physics. In quasi-1D free fermion systems at low temperature with quenched disorder, it has been established that in the diffusive (or metallic) regime the leading contribution to the large deviation function coincides exactly with that of the SSEP~\cite{Derrida2004}. Subleading quantum corrections, however, retain a clear imprint of the model’s {\it symmetry class} under time reversal. A well-known result due to Lee et al. states that \cite{LeeLevitovYuPRBUniversalstatistics}
\begin{equation}\label{eq:UCF}
{\rm var}\left(\frac{\left\langle  Q(t)\right\rangle }{t}\right)=\frac{2}{15\beta}\left(\frac{eV}{h}\right)^{2}
\end{equation}
where $V$ denotes the voltage, $e$ the electron charge, and $h$ Planck’s constant. The parameter $\beta$ is the {\it Dyson index}, taking the values $\beta=1,2,4$ for the known random matrix ensembles GOE, GUE, and GSE, respectively. In practice, one fixes a particular disorder realization, determines the mean transferred charge $\langle  Q(t)\rangle$ through repeated measurements, and then evaluates the variance over an ensemble of disorder configurations.
A straightforward computation shows that, {for the dynamical noise considered here, the quantity analogous to eq.~\eqref{eq:UCF} is given by} the second cumulant generated by $\delta F_{\rm qu}$ (the first vanishes, since the averaged dynamics is classical):
\begin{align}\label{eq:var-q-ucf}
-\frac{1}{N}\frac{\de^2}{\de u^2}\delta F_{{\rm qu}}(u)\Big\vert_{u=0} & \!\!\!\!\!\!\!\!\!=\frac{\mathbb{E}\left[\left\langle  Q(t)\right\rangle ^{2}\right]-\mathbb{E}\left[\left\langle  Q(t)\right\rangle \right]^{2}}{t}={\frac{{\rm var}\langle  Q(t)\rangle}{t}},
\end{align}
where the variance is taken with respect to the noise average $\E[\ ]$. In our case, the situation differs substantially, since the noise is explicitly time dependent. This feature complicates, among other aspects, the measurement of the universal conductance fluctuation in~\eqref{eq:UCF}, as it requires precise control over the dynamical noise. Nonetheless, {we mention that fine-grained control of time-dependent noise can be} achieved in quantum simulators and
is used, for instance, in random-circuit sampling protocols employed in demonstrations of quantum supremacy~\cite{Supremacy_Aaronson, Supremacy_Boixo,Supremacy_Bouland}. {Whether this degree of control is sufficient to measure current fluctuations for fixed noise realizations in the present setting lies beyond the scope of this work.}\\

We will show that a simple relation connects $F_{{\rm qu},w}(u)$ and $\delta F_{\rm qu}$ with cumulants of the {matrix of} the two–point function
\begin{equation}
G^{(w)}_{ji}(u) := \langle \hat{c}_j^\dagger \hat{c}_i\rangle_{w,u} ,
\end{equation}
where we introduced the notation 
\[
\langle \bullet \rangle_{w,u} := \frac{{\rm tr}(\hat{\rho}_{t,w}(u)\,\bullet)}{Z_{t,w}(u)}.
\]
In particular, to leading order in $1/N$, the calculation of $F_{\rm cl}(u)$ reduces to computing the one–point function
\begin{equation}\label{eq:g1-def}
\g_1(x=j/N) := \lim_{N\to\infty} \mathbb{E}\!\left[ G^{(w)}_{jj} \right](u),
\end{equation}
while the first subleading quantum correction in $\delta F_{\rm qu}$ is governed by
\begin{equation}\label{eq:g2-def}
\mathfrak{g}_{2}(x=j/N,\, y=i/N) := \lim_{N\to\infty} N \, \mathbb{E}\!\left[ G^{(w)}_{ji}\, G^{(w)}_{ij} \right]^c(u),
\end{equation}
where the superscript $c$ is used to denote connected correlations.\\

Both $\g_1(x)$ and $\g_2(x,y)$ are ${\cal O}(1)$ functions in the limit $N\to \infty$. We also remark that they depend on the bias parameter $u$, reducing to those found in Ref.~\cite{Bernard2019} when $u=0$. {Such definitions can be generalized to higher-order loops as
\be
\g_n(x_1,\dots, x_n) := \lim_{N\to\infty} N^{n-1} \, \mathbb{E}\!\left[ G^{(w)}_{i_1,i_2}\,  \dots  \, G^{(w)}_{i_ni_1}\right]^c(u),
\ee}
$x_k=i_k/N$.
\section{Exact calculations for the current statistics: general strategy}\label{sec:exact-calc}
We begin by observing that the biased quantum dynamics~\eqref{eq:dyn-bias} preserves the Gaussian character of the evolving state.  
For $u=0$ this follows immediately: the evolution is generated by a quadratic stochastic Hamiltonian~\eqref{eq:dH} together with boundary terms specified by the linear jump operators~\eqref{eq:Lbdry}.  The biased deformation~\eqref{eq:dH-bias} maintains this structure at all times; see Appendix~\ref{app:gaussian-prop} for a detailed discussion.  An important consequence is that the equations of motion for the two–point function $G^{(w)}_{ij}(u)$ form a closed system. They remain, however, nontrivial to solve, since the non-Hermitian character of the biased dynamics introduces intrinsic nonlinearities.
We also note that, although the noise–averaged evolution is not Gaussian-preserving, any averaged observable may still be obtained from the single–trajectory two–point function through Wick's theorem.\\

From the definition of the cumulant-generating function \eqref{eq:def-Fw}, one has
\be\label{eq:Ftmp}
\E[F_{{\rm qu},w}(u)] =  \lim_{N\to\infty} \frac{1}{N}\frac{d}{d\tau} \E[\log Z_{t,w}(u)],
\ee
{with $t=\tau N^2$.} A direct computation using the form \eqref{eq:dyn-bias} of the biased quantum evolution yields
\begin{align}\label{eq:Ftmp2-2}
\E[F_{{\rm qu},w}(u)] =N&\sum_{j=0}^{N-1}\Big\{\E\Big[ \left(e^{-u\lambda_j}-1\right)\langle \hat n_{j+1}\big(1+\sign \hat n_j\big)\rangle_{w,u}\nn
&+\left(e^{u\lambda_j}-1\right)\langle\hat n_{j}\big(1+\sign \hat n_{j+1}\big)\rangle_{w,u}\Big]\nn
&-4\sinh^2(u\lambda_j/2) \E[G^{(w)}_{j,j+1} G^{(w)}_{j+1,j}]^c(u)\Big\},
\end{align}
{with $\hat n_j:=\hat c^\dagger_j \hat c_j$}, and after using Wick's theorem
\begin{align}\label{eq:Ftmp2}
\E[F_{{\rm qu},w}(u)] =N&\sum_{j=0}^{N-1}\Big\{ \left(e^{-u\lambda_j}-1\right)\E[n^{(w)}_{j+1}](u)\big(1+\sign \E[n^{(w)}_j](u)\big)\nn
&+\left(e^{u\lambda_j}-1\right)\E[n^{(w)}_{j}](u)\big(1+\sign \E[n^{(w)}_{j+1}](u)\big)\nn
&+4\sign \sinh^2(u\lambda_j/2) \E[n^{(w)}_j n^{(w)}_{j+1}]^c(u)\Big\},
\end{align}
where $n_i^{(w)}(u):=G^{(w)}_{ii}(u)$. On the other hand, we have for the classical contribution 
\begin{align}
    F_{{\rm cl}}(u)=\underset{N\to\infty}{\lim} \frac{\log\mathbb{E}\left[Z_{t,w}\left(u\right)\right]}{N \tau},
    \label{eq:fullLDF}
\end{align}
and again, a direct computation leads to 
\begin{align}
F_{{\rm cl}}\left(u\right)=N\sum_{j=0}^{N-1}&\Big\{ \left(e^{-u\lambda_j}-1\right)\langle \hat n_{j+1}\big(1+\sign  \hat n_j\big)\rangle_{{\rm cl},u}\nn
&+\left(e^{u\lambda_j}-1\right)\langle \hat n_{j}\big(1+\sign \hat n_{j+1}\big)\rangle_{{\rm cl},u}\Big\},
\label{eq:classicalLDF}
\end{align}
where we introduced the ``classical" average \[
\langle \bullet \rangle _{{\rm cl},u}:=\frac{{\rm tr}\left(\bar{\rho}_{t}(u) \ \bullet\right)}{\mathbb{E}\left[Z_{t,w}(u)\right]}.\]

 The steady-state matrix $G^{(w)}_{ij}(u)$ in eq.~\eqref{eq:Ftmp2} is a representative of the ensembles of structured random matrices previously introduced in Refs.~\cite{Bernard2024,ExtraBH2025,Bernard2025} for the unbiased case $u= 0$, corresponding to the `open QSSEP' or `open QSSIP'.  This class of structured random matrices $M$ is characterized by the same properties (i.e. also valid for $u\neq 0$), \hypertarget{axioms}{which are}~\cite{Bernard2024,ExtraBH2025,Bernard2025}
\begin{itemize}
\item[{\it (i)}]~Local $U(1)$ invariance: in law, $M_{i,j} = e^{-i\theta_i} M_{i,j} e^{i\theta_j}$ for arbitrary phases $\theta_i$ and $\theta_j$. This condition follows from the $u=0$ case by inspecting the action of the bias on the quantum dynamics.

\item[{\it (ii)}]~Cumulants of cyclic products of matrix elements of order $n$ (or `loops') scale as $N^{1-n}$, i.e., $\E[M_{i_1,i_2} M_{i_2,i_3} \dots M_{i_n,i_1}]^c = \mathcal{O}(N^{1-n})$, with these cumulants continuous at coinciding indices.

\item[{\it (iii)}]~To leading order in $N$, the expectation values of disjoint loops factorize into the product of the expectation values of each loop; this property persists even in the limit of touching loops.

\item[{\it (iv)}]~Cumulants of $r$ disjoint loops, altogether comprising $n$ matrix entries, scale as $\mathcal{O}(N^{2-r-n})$, with the scaling in {\it (ii)} corresponding to the special case $r=1$.
\end{itemize}

The following relation between the two averages,
\be\label{eq:rel-avg}
\langle \ \rangle_{{\rm cl},u}=\E[\langle \ \rangle_{w,u}]\left(1+{\cal O}(N^{-2})\right),
\ee
holds, in law, for a generic product of number operators, as shown in Appendix~\ref{app:convergence}. {At $u=0$, eq.~\eqref{eq:rel-avg} is exact~\cite{Bernard2019}}.\\

{We recall that $\lambda_j\sim{\cal O}(1/N)$, as follows from \eqref{eq:gauge}.} Using this, together with eq.~\eqref{eq:rel-avg} and the axioms \hyperlink{axioms}{{\it(i-iv)}} listed above, one immediately concludes that, to leading order in $1/N$, the mean of the cumulant-generating function in \eqref{eq:Ftmp2-2} converges to its classical value~\eqref{eq:classicalLDF}, i.e. $\E[F_{{\rm qu},w}(u)]=F_{\rm cl}(u)+{\cal O}(N^{-1})$, as shown explicitly below. \\

Similarly, one finds for the leading quantum corrections to the classical behavior,
 \begin{align}
    \delta F_{{\rm qu}}(u)=&-4N\sum_{j=0}^{N-1}\sinh^{2}\left(u\lambda_{j}/2\right)\mathbb{E}\Big[G_{j,j+1}^{(w)}G_{j+1,j}^{(w)}\Big]^c\!\!(u)\nn
    &+{\cal O}(N^{-2}).
    \label{eq:quantumLDF}
\end{align}

{It is important to stress that the $1/N$ corrections to the cumulant-generating functions in eqs.~\eqref{eq:Ftmp2-2}–\eqref{eq:classicalLDF} have two qualitatively different origins. On the one hand, we find $1/N$ corrections arising from the next-to-leading order expansion of the biased density. We refer to these as {\it boundary corrections}, since they carry an explicit dependence on the specific choice of the boundary injection and extraction rates~\eqref{eq:Lbdry}~\cite{Derrida2004}. Since $\langle \hat n_i \rangle_{{\rm cl},u}=\E[n_i^{(w)}](u)\big(1+{\cal O}(N^{-2})\big)$, such boundary contributions are automatically subtracted in our definition~\eqref{eq:def-dFqu} of $\delta F_{\rm qu}$. The second type of $1/N$ corrections originates instead from the leading-order contribution of connected density--density correlations, and displays qualitative differences between the classical~\eqref{eq:classicalLDF} and quantum~\eqref{eq:Ftmp2-2} cases. Indeed, using~\eqref{eq:rel-avg} and the axioms~\hyperlink{axioms}{{\it(i--iv)}}, one finds in the classical case
\be
\langle \hat n_j \hat n_{j+1}\rangle^c_{{\rm cl},u}
=\sign\E\!\left[G_{j,j+1}^{(w)}G_{j+1,j}^{(w)}\right]^c + {\cal O}(N^{-2}),
\ee
whereas in the quantum case~\eqref{eq:Ftmp2-2} this contribution is canceled, already at the microscopic level, by the presence of the additional It\=o correction $-\frac12(dZ_{t,w}/Z_{t,w})^2$. One may therefore interpret $\delta F_{\rm qu}$ as encoding the absence of this type of $1/N$ corrections in the current cumulants generated by $\E[F_{{\rm qu},w}(u)]$. In contrast to the previous ones, we shall refer to these as {\it bulk corrections}, since they depend only on the imbalance of the thermodynamic densities at the endpoints, $\Delta \bar n := \bar n_0 - \bar n_N$, and not on the specific choice of the two reservoirs. For instance, the bulk correction to the current variance in the SSEP reads as $\frac{1}{3N}(\Delta \bar n)^2$ \cite{Derrida2004}, and can be directly computed starting from the unbiased connected density--density correlation, whose exact microscopic expression can be found e.g. in Ref.~\cite{Derrida2007}.
 }
\\

 Introducing rescaled weight fields
\be
 \xi(x=j/N) := N \lambda_j
 \ee 
such that $\int_0^1 dx\ \xi(x)=1$ (cf eq.~\eqref{eq:gauge}), we take the continuum limit of eqs.~\eqref{eq:Ftmp2} and \eqref{eq:quantumLDF} at large $N$. This yields to the results anticipated in the previous section, which are
\begin{align}\label{eq:F-general}
&F_{\rm cl}(u)\overset{N\to\infty}{\simeq} \!\!u\int_0^1 dx \, \Big[ -\xi(x)\,\partial_x \g_1(x)\nn
&\qquad\qquad+ u\,\xi^2(x)\g_1(x)\big(1+\sign \g_1(x)\big) \Big],\\[4pt]
\label{eq:Fq-general}
&\delta F_{{\rm qu}}(u)\overset{N\to\infty}{\simeq}-\frac{u^{2}}{N}\int_0^1 dx\ \xi^{2}\left(x\right)\mathfrak{g}_{2}\left(x,x\right).
\end{align}
We remark that, at this stage, the functions $\g_1(x)$ and $\g_2(x,x)$ appearing in \eqref{eq:F-general} and \eqref{eq:Fq-general} remain to be determined, while $\xi(x)$ can be arbitrarily chosen such that its integral is one, for instance $\xi(x)\equiv 1$. \\

One may observe that $\delta F_{\rm qu}(u)$ involves, up to ${\cal O}(u^2)$, only unbiased $\g_2(x,x)$ correlation. This has already been determined in earlier works~\cite{Bernard2019}
\begin{equation}\label{eq:g2-u=0}
    \g_2(x,y)\Big\vert_{u=0} = (\bar{n}_0-\bar{n}_N)^2\, \big[ \min(x,y)-xy\big],
\end{equation}
and directly provides the universal conductance fluctuation (see eq.~\eqref{eq:UCF}) for the QSSIP/QSSEP
\begin{align}\label{eq:res-ucf}
    \frac{\de^2}{\de u^2}\,\delta F_{\rm qu}(u) \Big\vert_{u=0}
    = -\frac{(\bar{n}_0-\bar{n}_N)^2}{3N}.
\end{align}
Remarkably, this result is independent of the underlying bosonic or fermionic statistics. {This can be understood from the fact that the correction is determined entirely by the unbiased function $\g_2(x,x)$. Its expression \eqref{eq:g2-u=0} follows only from diffusion and particle-number conservation, which are the same for bosons and fermions~\cite{Bernard2025}. The correction therefore depends only on the difference between the boundary densities and not on the particle statistics.} The overall minus sign carries a conceptual significance: as anticipated, a careful comparison of \eqref{eq:res-ucf} with known expressions of the current variance in the SSEP \cite{Derrida2004,Bodineau2007}~reveals that this bulk contribution is absent in the quantum process. The same result can be obtained from a direct calculation relying solely on local charge conservation in the model, see Appendix~\ref{app:direct-calc-var}.\\

To generate higher–order contributions from~\eqref{eq:F-general}–\eqref{eq:Fq-general}, a full characterization of the biased correlations $\g_1(x)$ and $\g_2(x,x)$ is required. We begin with $\g_1(x)$, considering the evolution equation of the particle density. In the long-time limit, the latter is expected to converge to a stationary value, namely,
\be
\lim_{t\to\infty}\frac{d}{dt} \E\left[\frac{\tr(\hat\rho_{t,w}(u) \ \hat{n}_j)}{Z_{t,w}(u)}\right]=0.
\ee
Using It\={o} calculus, it is easy to see that the previous condition implies
\be\label{eq:stat}
\lim_{t\to\infty} \E\Bigg[\left(\frac{\tr\big(d\hat\rho_{t,w} \hat{n}_j\big)}{Z_{t,w}} 
       - \frac{\tr(\hat\rho_{t,w} \hat{n}_j)}{Z_{t,w}} \frac{dZ_{t,w}}{Z_{t,w}}\right) \left(1- \frac{dZ_{t,w}}{Z_{t,w}}\right)\Bigg] = 0.
\ee
After a lengthy but straightforward calculation based on the application of Wick's theorem and of the axioms~\hyperlink{axioms}{\it(i-iv)}, to leading order in $1/N$, eq.~\eqref{eq:stat} becomes
\begin{align}\label{eq:firstorderg}
&\de_x^2 \g_1-2u\xi(x)\de_x\g_1 \left[1+2\sign \g_1(x)\right]-u\g_1(x) \de_x\xi \left[1+\sign\g_1(x)\right]\nn
&\quad +u^2\xi^2(x)\g_1(x)\left[1+3\sign\g_1(x)\right]+2u^2\xi^2(x)\g_1^3(x)\nn
&=-\Lambda[\g_2](x),
\end{align}
where we introduced the notation
\begin{align}
\Lambda\left[f\right](x,\cdots):=u\int_0^1 \!\!dy&\left[u\xi(y)^{2}\left(2\g_{1}(y)+\sign\right)+\sign \partial_{y}\xi(y)\right]\nn
&\times f\left(x,\cdots,y\right)\! .
\end{align}
We observe that, for a general choice of $\xi(x)$, the function $\g_1(x)$ is determined self-consistently through higher–order loop contributions. The resulting hierarchy takes the form of a triangular system, which can be solved perturbatively in a small-$u$ expansion, as discussed in Appendix~\ref{app:low-order}. An alternative route opens if the term $\Lambda[g_2]$ is absent, in which case the hierarchy closes. As shown in Ref.~\cite{costa2025}, one may exploit the gauge freedom to enforce precisely this cancellation
\be\label{eq:gaugecondition}
\partial_x \xi(x) = -u\,\xi^2(x)\big(1+\sign 2\g_1(x)\big).
\ee
Together with \eqref{eq:firstorderg}, this yields the closed set of equations
\be\label{eq:cond1}\begin{cases}
\xi(x)\,\partial_x^2 \g_1(x) + 2\,\partial_x \g_1(x)\,\partial_x \xi(x) \\ \quad - 2u\, \g_1(x)\big(1+\sign \g_1(x)\big)\xi(x)\,\partial_x \xi(x) = 0, \\[10pt]
\partial_x \xi(x) = -u\,\xi^2(x)\big(1+\sign 2\g_1(x)\big);
\end{cases}
\ee
that can be solved with boundary conditions $\g_1(0) =\bar{n}_0$ and $\g_1(1) = \bar{n}_N$, as fixed by eq.~\eqref{eq:baths-dens-values} (since the bias 
does not act on the boundaries) and $\int_0^1 dx\ \xi(x)=1$.\\

The equation for $\mathfrak{g}_2(x,y)$ is obtained in the same fashion, starting from the stationarity condition  
\[
\lim_{t\to\infty}\frac{d}{dt}\,\E\!\left[ G^{(w)}_{ij} G^{(w)}_{ji} \right]^{\!c} = 0
\]
and applying It\=o calculus. After some algebra, and retaining only the leading terms in $1/N$, one finds  
\begin{align}\label{eq:secondorder_g}
 & \Lambda\left[{\mathfrak{g}^{\rm nc}_{3}}\right](x,y)+\int_0^1 dz\ u^{2}\xi^{2}(z)\left(\mathfrak{g}_{2}^{\rm nc}(x,z)\mathfrak{g}_{2}^{\rm nc}(z,y)\right)+\partial_{x}^{2}\mathfrak{g}_{2}^{\rm nc}(x,y)\nn
 &-\sign\mathscr{L}[\g_{2}^{\rm nc}](x,y)\left(2\g_{1}(x)+\sign\right)-\sign 2\mathscr{L}[\g_{1}](x)\mathfrak{g}_{2}^{\rm nc}(x,y)\nn
 &+\partial_x \partial_y (\delta(x,y)\g_{1}(x)\g_{1}(y))
  -\delta(x,y)\mathscr{L}\left[\mathfrak{g}_{1}\right]\left(x\right)\mathfrak{g}_{1}(x)\nn
  & +(\text{  \ensuremath{x\leftrightarrow y} })=0,
\end{align}
where 
\be
\mathscr{L}\left[f\right](x,\cdots):=\Big(u\left(\partial_{x}\xi(x)+2\xi(x)\partial_{x}\right)-u^{2}\xi^{2}(x)\Big)f\left(x,\cdots\right).
\ee
For convenience, we introduced the nonconnected two-point function defined as 
\begin{equation}
    \mathfrak{g}_2^{\rm nc}(x,y):=\mathfrak{g}_2(x,y)+\delta(x,y)\mathfrak{g}_1(x)\mathfrak{g}_1(y),
\end{equation}
{and similarly for nonconnected version of $\g_3$.} Once more, the term $\Lambda[\g_3^{\rm nc}]$ is the source of the non-closure of the stationary equations. Consequently, the {\it same} gauge condition of eq.~\eqref{eq:gaugecondition} enforces closure at this level. In fact, one can show that this property holds to all orders: the unique gauge choice \eqref{eq:gaugecondition} ensures closure of the stationary hierarchy at every order (see Appendix~\ref{app:gn-general} for a proof).\\

Even though the resulting equations are closed, they do not necessarily admit closed-form analytic solutions. For the first–order equation involving $\g_1(x)$, an explicit solution can indeed be obtained, as shown in Appendix~\ref{app:sol-gauge-eqs} and discussed below. For $\mathfrak{g}_2(x,x)$, however, we were unable to find an {explicit} solution. Nevertheless, in Sec.~\ref{sec:quant-fluc} we present a perturbative calculation based on small-$u$ expansion.

\section{Classical fluctuations: results and equivalence with MFT}\label{sec:classical}
We first begin by discussing the classical fluctuations encoded in $\mathfrak{g}_1(x)$. Eqs.~\eqref{eq:cond1} admit a closed-form solution (see Appendix~\ref{app:sol-gauge-eqs} for details), leading to an explicit form for the cumulant-generating function,
\be\label{eq:F-exact}
\lim_{N\to\infty} \E[F_{{\rm qu},w}(u)]=\begin{cases}\displaystyle
-\sign\left(\cosh^{-1}\sqrt{\omega(u)}\right)^2; \quad \omega(u)\geq 1\\[10pt]\displaystyle
\sign\left(\cos^{-1}\sqrt{\omega(u)}\right)^2;\quad \omega(u)<1
\end{cases}
\ee
with
\be
\omega(u):=\big[1-\sign \bar{n}_0(e^{u}-1)\big]\big[1-\sign \bar{n}_N(e^{-u}-1)\big].
\ee

In the fermionic case ($\sign=-1$), $\E[F_{{\rm qu},w}(u)]$ reduces to the well-known result for the SSEP~\cite{Derrida2004,Derrida2007}, highlighting the classical character of the  transport properties in the large-$N$ limit. To the best of our knowledge, the expression for the bosonic case ($\sign=+1$), corresponding to the classical inclusion process, has not appeared previously in the literature. As for the large deviation of density fluctuations~\cite{Bernard2025}, the bosonic and fermionic expressions differ only by a few signs, which are tracked throughout the calculation with the parameter $\sign$. \\

Beyond this similarity, however, we find a qualitative difference in the SSIP. For $\sign=+1$, the cumulant-generating function $\E[F_{{\rm qu},w}(u)]$ in eq.~\eqref{eq:F-exact} is defined only for $u\in(u_{-},u_{+})$, with
\be\label{eq:threshold}
u_{+} = \log\!\left(1+\frac{1}{\bar{n}_0}\right), \quad u_{-} = -\log\!\left(1+\frac{1}{\bar{n}_N}\right).
\ee
This restriction reflects the fact that, in the bosonic case, large positive (negative) values of $u$ bias the dynamics towards arbitrarily high injection (extraction) currents, which cannot be sustained: instead of producing unbounded currents, the system undergoes boundary condensation at the leftmost (rightmost) site. This can be seen already at the single-site level. The probability $P_0$ to have $n$ particles at site $0$ follows the geometric law
\be
P_0(n)\propto z^n,\qquad z=\frac{\bar{n}_0}{1+\bar{n}_0}.
\ee
Biasing the dynamics amounts to replacing the fugacity $z\mapsto z e^u$, so that the local partition sum reads
\be\label{eq:MGF-2}
Z(u)=\sum_{n=0}^\infty (z e^u)^n,
\ee
which is finite only if $z e^u<1$, i.e., $u<u_+$.  It follows that the biased occupation on the leftmost site, $\sum_{n=0}^\infty n \, (z e^u)^n$, diverges for $u > u_+$, leading to boundary-driven condensation in the presence of strong bias. Hence, the threshold \eqref{eq:threshold} originates already from the (noninteracting) boundary problem and is independent of the bulk dynamics. In general, the bulk dynamics and the right boundary in SSIP do affect the current fluctuations, but they cannot suppress this divergence. The same reasoning applies for large negative bias, leading to $u<u_-$. Expanding around the closest singularity $u_0 \in \{u_+,u_-\}$ and applying Darboux's theorem, the large-order asymptotics of the currents cumulants reads
\be
\frac{\de^q}{N\de u^q} \E[F_{{\rm qu},w}(u)]\Big\vert_{u=0} \sim \frac{\sqrt{\pi}}{2N}\,\sqrt{|u_0|} \sqrt{\bar n_0+\bar n_N+1}\; \frac{q!}{|u_0|^q\, q^{3/2}},
\ee
with an additional factor $(-1)^q$ if the dominant singularity lies at $u_-<0$. Thus, cumulants of the QSSIP grow factorially, with exponential scale set by the nearest branch point and with $q^{-3/2}$ algebraic prefactor.\\

In contrast, in the SSEP ($\sign=-1$) the local occupation is bounded ($n=0,1$), so eq.~\eqref{eq:MGF-2} is always finite and the generating function remains well defined for all $u$, allowing the system to sustain arbitrarily large positive or negative currents.\\

\begin{figure}[t]
\centering
\includegraphics[width=.95\columnwidth]{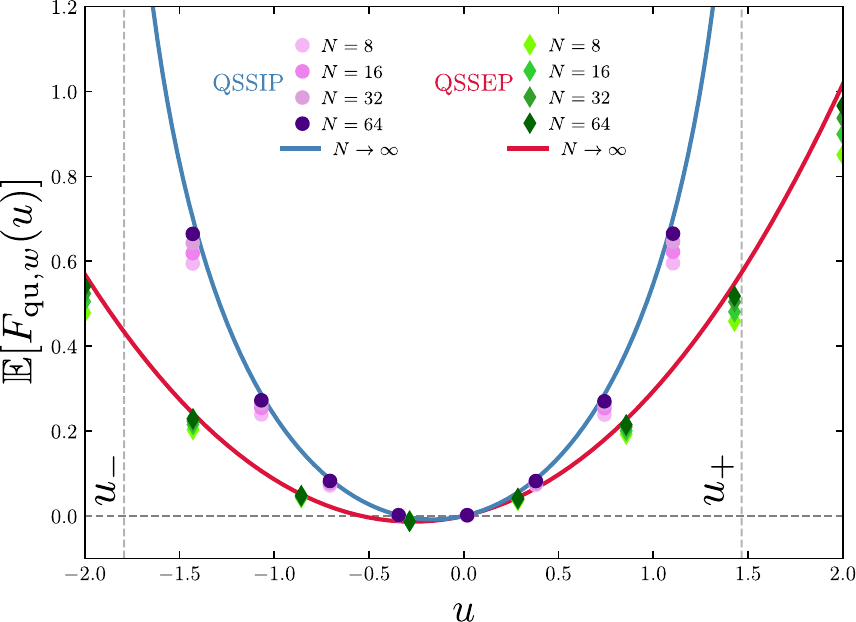}
\caption{Plot of the cumulant-generating function $\E[F_{{\rm qu},w}(u)]$ in \eqref{eq:F-exact} (solid lines) for the bosonic and fermionic case ($\sign=\pm 1$, respectively). Symbols show the numerical data for the mean from the microscopic dynamics~\eqref{eq:dyn-bias} for different system sizes $N\leq 64$ (see plot's legend). Left- and right- reservoir densities are set to $\bar{n}_0=0.3$ and $\bar{n}_N=0.2$. In the bosonic case, the function $\E[F_{{\rm qu},w}(u)]$ is defined in the interval $u\in(u_-,u_+)$ (dashed vertical lines).}\label{fig:F}
\end{figure}
\begin{figure}[t]
\centering
\includegraphics[width=.95\columnwidth]{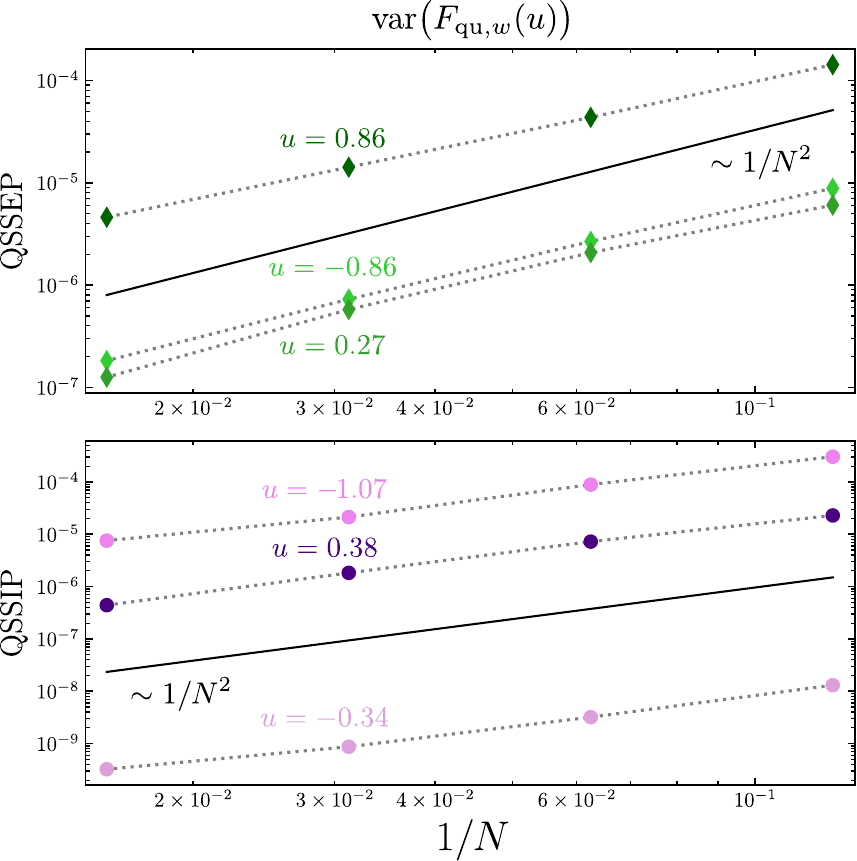}
\caption{Numerical analysis of the variance $\text{var}\big(F_{{\rm qu},w}(u)\big):=\E[F_{{\rm qu},w}(u)^2]-\E[F_{{\rm qu},w}(u)]^2$, 
for the fermionic (top) and bosonic (bottom) cases, shown as a function of $1/N$ 
for different values of the bias amplitude $u$. The data are compatible with a 
$\sim 1/N^2$ scaling of the variance, indicated by the solid line as a guideline. Left- and right- reservoir densities are set to $\bar{n}_0=0.3$ and $\bar{n}_N=0.2$.
}\label{fig:variance}
\end{figure}

We also notice that $\E[F_{{\rm qu},w}(u)]$ satisfies the following symmetries~\cite{Derrida2004}:
\begin{itemize}
\item[-] {\it Left-right symmetry}~--~if the left and right reservoirs are exchanged (i.e., $\g_1(0)= \bar{n}_N$ and $\g_1(1)= \bar{n}_0$), the current statistics remain unchanged under $u\to -u$.  
\item[-] {\it Gallavotti-Cohen symmetry}~--~the probabilities of currents from left to right and from right to left are exponentially related, reflecting a fluctuation symmetry in the nonequilibrium steady state. Explicitly, $\E[F_{{\rm qu},w}(u)]=\E[F_{{\rm qu},w}(z-u)]$ with 
\[
z=\log\left[\frac{\bar{n}_N\big(1+\sign \bar{n}_0\big)}{\bar{n}_0\big(1+\sign \bar{n}_N\big)}\right].
\]  
\end{itemize}
For the fermionic case ($\sign=-1$), i.e. for the classical exclusion process, one further has
\begin{itemize}
\item[-] {\it Particle-hole symmetry} ~--~exchanging particles and vacancies, $\bar{n}_0\to 1-\bar{n}_0$ and $\bar{n}_N\to 1-\bar{n}_N$, leaves the fermionic current statistics invariant under $u\to -u$. This symmetry does not hold in the bosonic case.
\end{itemize}

Figure~\ref{fig:F} shows the cumulant-generating function for the two cases. Numerical data for $\E[F_{{\rm qu},w}(u)]$ from the microscopic dynamics~\eqref{eq:dyn-bias} for modest system sizes ($N\leq 64$) up to times $t={\cal O}(N^2)$ show mean convergence toward the results in \eqref{eq:F-exact}. Such convergence is nonuniform in the bias amplitude $u$. See Appendix~\ref{app:numerics} for details on the numerical implementation. Our numerical analysis of the variance of $F_{{\rm qu},w}(u)$ in Fig.~\ref{fig:variance} indicates the scaling
\be
F_{{\rm qu},w}(u) \overset{N\to\infty}{\simeq} F_\text{lead}(u) + \frac{f(u)}{N^\upsilon},
\ee
{where $F_\text{lead}\sim {\cal O}(1)$ denotes the leading contribution of $\E[F_{{\rm qu},w}]\simeq F_{\rm cl}$ in eq.~\eqref{eq:F-general}}, and $f(u) \sim {\cal O}(1)$ with $\upsilon \simeq 1$ according to our data. This behavior is consistent with {the self-averaging estimate derived in Sec.~\ref{sec:var} below and with} the expected scaling of the subleading boundary corrections; see the discussion below eq.~\eqref{eq:quantumLDF}. Moreover, fluctuations of the integrated current decay faster with system size than local density fluctuations, which scale as $1/\sqrt{N}$ \cite{Bernard2025}. A more systematic analytical analysis, complemented by a broader numerical exploration, would be required to fully substantiate these observations.

\subsection{Equivalence with MFT}\label{sec:classical-MFT}
In this section we show that the equations~\eqref{eq:cond1} determining $\E[F_{{\rm qu},w}(u)]$ to leading order in $1/N$ are analogous to the MFT equations derived for the corresponding classical models. One starting point in MFT is to assume that the density $\n(x,\tau)$ and the current $\jj(x,\tau)$ satisfy the stochastic equations~\cite{bertini2001fluctuations,bertini2002macroscopic}
\begin{align}
&\de_\tau \n(x,\tau) + \de_x \jj(x,\tau) = 0; \label{eq:cons-law}\\
&\jj(x,\tau) = - \de_x \n(x,\tau) + N^{-1/2} \, \sqrt{\sigma(\n)} \, \eta(x,\tau), \label{eq:fick}
\end{align}
where $\tau = t/N^2 \sim \mathcal{O}(1)$ is the rescaled time at diffusive scales, and $\sigma(\n) = 2 \n (1 + \sign \n)$ is the mobility of the SSIP or SSEP for $\sign = \pm 1$, respectively. The diffusion constant of the underlying models is $D = 1$, and $\eta(x,\tau)$ is a Gaussian space-time white noise with
\begin{equation}
\E[\eta(x,\tau)\, \eta(x',\tau')] = \delta(x-x') \, \delta(\tau-\tau').
\end{equation}
Equation~\eqref{eq:cons-law} expresses a conservation law, while~\eqref{eq:fick} is a noisy version of Fick's law. By integrating out the noise, one can write the joint probability of observing a density $\n(x,\tau)$ and current $\jj(x,\tau)$ profile in the system as
\be
\text{Prob}(\n, \jj) \propto e^{-NS[\n,\jj]}, \quad S[\n,\jj]:= \int_0^\tau ds\ \int_0^1 dx \ \frac{\big(\jj + \de_x\n\big)^2}{2\sigma(\n)},
\ee
conditioned by the conservation law \eqref{eq:cons-law}. One can then introduce the current moment-generating function as
\be\label{eq:Z-mft}
Z_t(u):=\E\left[\exp\left(uN\int_0^\tau ds \int_0^1 dx  \ \jj(x,s)\right)\right],
\ee
and rewrite the corresponding MFT action in terms of the fields $\n$ and the response field $\p$, namely
\be\label{eq:Smft}
S_u[\n,\p]=\int_0^\tau ds \int_0^1 dx \Big(-\p\,\de_s \n + \mathscr{H}_u[\n,\p] \Big) + S^\text{bdry}_u,
\ee
with definition of the biased current $\jj_u := -\de_x\n + \sigma(\n)(\de_x\p-u)$, and we introduced the Hamiltonian density
\be\label{eq:hamiltonian-mft}
\mathscr{H}_u[\n,\p]=\n \de_x^2 \p +\frac{\sigma(\n)}{2}\big(\de_x\p-u\big)^2.
\ee
The boundary term reads as
\be\label{eq:Smft-bdry}
S^\text{bdry}_u[\n,\p] = -\int_0^\tau ds \; \Big[\, \p \jj_u +  \n \, (\partial_x \p-u) \,\Big]_{x=0}^{x=1}.
\ee
 It follows that the long-time limit of the moment-generating function is controlled by
\be
Z_t(u) \underset{N\to \infty}{\asymp} e^{N\tau \int dx\, \mathscr{H}_u[\n_*,\p_*]},
\ee
and fields $\n_*,\p_*$ determined by the following steady-state Hamilton equations of the corresponding MFT action \eqref{eq:Smft} (also referred to as `{stationary} MFT equations')~\cite{Derrida2011}
\be\label{eq:MFT-eqs}
\begin{cases}\displaystyle
\frac{\delta \mathscr{H}[\n,\p]}{\delta \p}=0 = \de_x^2 \n - \de_x\!\left[\sigma(\n)\big(\de_x\p-u\big)\right],\\[12pt]\displaystyle
\frac{\delta \mathscr{H}[\n,\p]}{\delta \n}=0= \de_x^2\p + \tfrac12 \sigma'(\n)\, \big(\de_x\p-u\big)^2.
\end{cases}
\ee
By setting
\be\label{eq:relation-xi-response}
 \de_x\p-u = u\xi(x); \quad \n(x) = \g_1(x),
\ee
one can easily show that the MFT equations~\eqref{eq:MFT-eqs} are equivalent to the extremization conditions in eq.~\eqref{eq:cond1}. The boundary values of the fields $\n$ and $\p$ are determined from the boundary term arising in the action \eqref{eq:Smft} under variations of the fields $\n$ and $\p$. Because the field $\n(x)$ satisfies Dirichlet boundary conditions, see eq.~\eqref{eq:baths-dens-values}, we need to vary \eqref{eq:Smft}-\eqref{eq:Smft-bdry} only with respect to the response field $\p$. 
This gives
\be\label{eq:bdry-var}
\delta_\p S_u[\n,\p] = - \int_0^\tau ds \; \jj_u(x,s) \, \delta \p(x,s) \Big\vert_{x=0}^{x=1},
\ee
where $\jj_u$ is the biased current defined above. Since, in the open setup, the biased current $\jj_u(x,s)$ does not vanish at the boundaries,  the boundary variation \eqref{eq:bdry-var} enforces Dirichlet boundaries also for the field $\p(x)$. By integrating \eqref{eq:relation-xi-response}, one finds
\be
\p(1)-\p(0)=u + u \int_0^1 dx\ \xi(x).
\ee
As the response field $\p(x)$ couples the density only through derivative terms in \eqref{eq:hamiltonian-mft}, a convenient gauge is $\p(1)-\p(0)=2u$. This choice yields the normalization $\int_0^1 dx\, \xi(x) = 1$, in agreement with the result obtained from the exact calculation of Sec.~\ref{sec:exact-calc}.

\subsection{Classical typicality {and} variational approach}\label{sec:var}
Motivated by the correspondence between the leading current fluctuations and the variational framework supplied by MFT, we now examine the quantum problem {and address the classical typicality of the current statistics at the level of a single stochastic realization}. We start from the noise-averaged (classical) process and consider the following eigenbasis of ${\cal L}_{u}$,
\be\label{eq:eigenbasis}
{\cal L}_{u}(\hat{L}_{n}) = \Psi_{n}(u) \hat{L}_{n},\quad
{\cal L}^\dagger_{u}(\hat{R}_{n}) = \overline\Psi_{n}(u) \hat{R}_{n},  
\ee
with $\tr(\hat{L}_{n} \hat{R}_{m}) = \delta_{nm}$. {Here and below, ${\cal L}_u$ denotes the full tilted generator, including ${\cal L}_{\rm bdry}$.} In this basis, the noise-averaged biased density matrix can be written as
\be
\bar\rho_t(u) =\sum_n  \tr(\bar\rho_0 \hat{R}_n)\ e^{t\Psi_n(u)} \hat{L}_n
\ee
since $\bar\rho_t(u)$ acts as a left operator in the Hilbert-Schmidt scalar product sense. Assuming that the spectrum of ${\cal L}_{u}$ has a gap, i.e., $D_{n}(u) := \Psi_{n}(u) - \Psi_{0}(u)$, with ${\rm Re}[D_{n}(u)] < 0$, the long-time behavior of $\tr\big(\hat\rho_t(u)\big)$ is dominated by the principal eigenvalue $\Psi_{0}(u)$,  
\be\label{eq:growth-mean}
\tr\big(\bar\rho_t(u)\big)=\E[Z_{t,w}(u)]\underset{t\to\infty}{\asymp} e^{t \Psi_0(u)}.
\ee
The cumulant-generating function for the classical process can therefore be identified with the long-time growth rate in~\eqref{eq:growth-mean}, $F_\text{cl}(u)=N{\rm Re}\Psi_0(u)$. Its value can be obtained from the variational formula  
\be\label{eq:var-mean}
\frac{F_\text{cl}(u)}{N} = {\rm Re}\,\Psi_{0}(u) =  {\rm Re}\, \underset{\hat{L}, \hat{R} >0}{\text{extr}} \,  \frac{\tr\big(\hat{L}  {\cal L}_{-u}(\hat{R})\big)}{\tr\big(\hat{L}\, \hat{R}\big)}.
\ee

\indent
We now turn to the stochastic quantum process. In the {left/right} eigenbasis~\eqref{eq:eigenbasis} of ${\cal L}_{u}$, the single-realization density matrix can be expanded as
\be
\hat\rho_{t,w}(u) = \sum_n c^{(n)}_{t,w}(u)\, e^{t \Psi_{n}(u)} \hat{L}_{n},
\ee
where the coefficients satisfy the linear stochastic equations
\be\label{eq:coef}
dc^{(n)}_{t,w}(u)=\sum_m \tr\!\Big(\hat R_n\,d{\cal H}_{w,u}(\hat L_m)\Big) e^{t(\Psi_m(u)-\Psi_n(u))}c^{(m)}_{t,w}(u).
\ee
This defines a stochastic problem whose long-time behavior is governed by a Lyapunov exponent, controlling the asymptotic growth of $\hat\rho_{t \to \infty, w}(u)$. {
Despite its linear form, eq.~\eqref{eq:coef} describes a nontrivial stochastic mode-mixing problem: all modes are coupled by the noise, and the deterministic factors $e^{t(\Psi_m-\Psi_n)}$ alone do not imply the decay of the nonprincipal contributions. A direct evaluation of the corresponding Lyapunov exponent is therefore difficult and will not be needed here. Instead, we introduce the density matrix normalized by the noise-averaged growth,
\be\label{eq:normalized-state}
\hat\Delta_t:=e^{-t\Psi_0(u)}\hat\rho_{t,w}(u).
\ee
Since $Z_{t,w}(u)=e^{t\Psi_0(u)}\tr(\hat\Delta_t)$, at the diffusive time $t=\tau N^2$ one has
\be\label{eq:typicality-identity}
F_{{\rm qu},w}(u)-F_{\rm cl}(u)=\lim_{N\to\infty} \frac{X_{\tau N^2}}{\tau N},
\ee
with $X_t:=\log\tr(\hat\Delta_t)$. We argue that $X_{\tau N^2}$, which controls the right-hand side of eq.~\eqref{eq:typicality-identity}, is bounded in both mean and variance. The bound on the mean follows from the exact calculation of Sec.~\ref{sec:exact-calc} and gives the mean convergence to the classical result established above. A bound on the variance further shows that fluctuations around this mean vanish as $N\to\infty$, thus extending the previous result to a typical realization of the noise. This variance bound can be obtained by inspecting the stochastic evolution of $X_t$, as shown in Appendix~\ref{app:details-var}. The proof requires the additional natural assumption that traceful modes which contribute to the cumulants at diffusive times $t=\tau N^2$ have diffusive gaps $D_n={\cal O}(N^{-2})$, and that summing their contributions does not overcome this $N^{-2}$ scaling; see also Refs.~\cite{nidari2014b,nidari2015}. This assumption applies only to the modes that contribute at diffusive scales; fast or traceless modes do not need to satisfy it.
Under this assumption, for fixed $u$ and $\tau$, one finds (see Appendix~\ref{app:details-var} for a derivation)
\be\label{eq:normalized-log-bounds}
\E[X_{\tau N^2}]={\cal O}(1), \quad \text{var}(X_{\tau N^2})={\cal O}(1).
\ee
Using eq.~\eqref{eq:typicality-identity}, we then obtain
\be\label{eq:var-bound}
\E\big[\big(F_{{\rm qu},w}(u)-F_{\rm cl}(u)\big)^2\big]={\cal O}(N^{-2}).
\ee
Thus, the single-realization cumulant-generating function $F_{{\rm qu},w}$ is self-averaging.} This argument supports the conjectured classical typicality of the quantum process in the large-$N$ limit
\begin{equation}
F_{{\rm qu},w}(u) \,\overset{N\to\infty}{\sim}\, F_\text{cl}(u).
\end{equation}

{Combining the self-averaging result with the classical variational formula~\eqref{eq:var-mean}, the leading typical single-realization growth rate can be written as
\be\label{eq:F-var}
F_{{\rm qu},w}(u)=N\,{\rm Re}\,\underset{\hat L,\hat R>0}{\mathrm{extr}}\,{\cal F}_u(\hat L,\hat R)+{\cal O}(N^{-1}),
\ee
where
\be\label{eq:var-func-def}
{\cal F}_u(\hat L,\hat R):=\frac{\tr\big(\hat L\,{\cal L}_{-u}(\hat R)\big)}{\tr\big(\hat L\,\hat R\big)}.
\ee}
Given that the biased stochastic dynamics~\eqref{eq:dyn-bias} preserves Gaussianity for each realization, it is natural to evaluate this functional on Gaussian trial states of the form
\be\label{eq:trial-states}
\hat L,\hat R= \exp\bigg(\sum_{i,j}K^{(L/R)}_{ij}\, \hat c_i^\dagger\hat c_j\bigg).
\ee
A direct computation allows one to evaluate ${\cal F}_u$ for arbitrary Gaussian states; see Appendix~\ref{app:details-extr-variational-app}.\\

{Remarkably, the extremization conditions
\be\label{eq:var-ansatz-extr}
 \frac{\delta}{\delta \hat R} {\cal F}_u\Big\vert_{\hat L_0, \hat R_0}=0,
 \qquad
 \frac{\delta}{\delta \hat L} {\cal F}_u\Big\vert_{\hat L_0, \hat R_0}=0
\ee
admit a {\it diagonal} Gaussian solution and yield, to leading order in $1/N$, the same closed set of equations as Eq.~\eqref{eq:cond1}. At this extremum, 
\be
N\,{\rm Re}\,{\cal F}_u(\hat L_0,\hat R_0)=F_{\rm cl}(u)+{\cal O}(N^{-1}).
\ee}
We refer to Appendix~\ref{app:details-extr-variational-app} for the derivation.\\

{Thus, the variational calculation provides a diagonal Gaussian representation of the leading classical value. Self-averaging explains this result. Although the noise-averaged dynamics is, in general, non-Gaussian, eq.~\eqref{eq:var-bound} shows that the growth rate of a typical realization concentrates on the noise-averaged value. Since each stochastic realization preserves Gaussianity, this common leading value can be extracted from a diagonal Gaussian manifold.}
\section{Absence of bulk corrections in the QSSIP and QSSEP cumulants}\label{sec:quant-fluc}
{Finally, we argue that the absence of $1/N$ bulk contributions, observed in the current variance (cf.~Eq.~\eqref{eq:res-ucf}), is expected to extend to higher-order cumulants. Indeed, as discussed in Sec.~\ref{sec:exact-calc}, the exact microscopic expression for $\E[F_{{\rm qu},w}(u)]$ does not contain the term originating from the leading ${\cal O}(1/N)$ behavior of connected density--density correlations, which would otherwise generate such bulk contributions to the current cumulants. All other subleading corrections in $\E[F_{{\rm qu},w}(u)]$ either depend on the specific choice of the baths (boundary-type contributions) or enter the cumulants only at order ${\cal O}(N^{-2})$. At present, however, a direct verification of this expectation is not possible due to the lack of explicit results for the corresponding bulk finite-size corrections in the classical current cumulants beyond the variance. On the quantum side, by contrast, these bulk contributions can be directly related to the biased scaling function $\g_2(x,y)$ (cf.~Eq.~\eqref{eq:quantumLDF}), and thus systematically computed.\\
} 

 Since an exact solution of the scaling function $\g_2(x,y)$ is not available, we present a small-$u$ perturbative solution of the stationary equation \eqref{eq:secondorder_g}. See also Appendix~\ref{app:low-order} for the perturbative calculation of $\g_1(x)$. Throughout this analysis we fix the gauge field to $\xi(x)\equiv 1$. For clarity, we restrict to the sector $x<y$; the complementary case $x>y$ follows by the same steps. Because the bias transformation leaves the boundaries unchanged, the function $\mathfrak{g}_{2}(x,y)$ satisfies  
\begin{equation}
\mathfrak{g}_{2}(0,y)=\mathfrak{g}_{2}(x,1)=0.
\end{equation}
The boundary condition at $x=y$ is instead fixed by the contact terms in \eqref{eq:secondorder_g},
\begin{align}
& \left(\partial_x - \partial_y\right) \g_2(x,y) \big|_{x=y^-} \nonumber \\
   = & - \int \left[\varepsilon u\partial_s \xi(s) + u^2 \xi^2(s)(2\g_1(s)+\varepsilon)\right]\g_2(x,s)  \ ds \ \g_1(x)  \nonumber\\
    &+ (\partial_x \g_1(x))^2 -\varepsilon u\partial_x \xi(x)\g_1^3(x)\nonumber \\
    & -2\varepsilon u \xi(x)\partial_x \g_1(x)\g_1(x) \left(\g_1(x)+\varepsilon\right) \nonumber \\
    &+ u^2 \xi^2(x)(2\g_1(x)+\varepsilon)(-\g_1(x)+\varepsilon) \g_1^2(x).     
\end{align}
We thus proceed by expanding in powers of $u$,
\begin{align}
&\g_1(x):= \g_1^{\{0\}}(x)+{\cal O}(u);\\
&\mathfrak{g}_{2}(x,y)=:\mathfrak{g}_{2}^{\{0\}}(x,y)+u \mathfrak{g}_{2}^{\{1\}}(x,y)+{\cal O}(u^2),
\end{align}
with $\g_1^{\{0\}}(x)=\bar{n}_0(1-x)+x \bar{n}_N$ as obtained in \cite{Bernard2019}, and with $\g_2^{\{0\}}$ corresponding to the unbiased function given in \eqref{eq:g2-u=0}. Starting from \eqref{eq:secondorder_g}, one obtains the following equation for $\g_2^{\{1\}}$ in the bulk whenever $x\neq y$:
\begin{align}
    \Delta \g_2^{\{1\}}(x,y) = & 4\sign \partial_x\left( \g_2^{\{0\}}(x,y) \g_1^{\{0\}}(x)\right) \nn
    &+ 2 \partial_x \g_2^{\{0\}}(x,y) +(\text{\ensuremath{x\leftrightarrow y} }).
\end{align}
Evaluating $\eqref{eq:secondorder_g}$ at the boundary $x=y-0^+$ yields 
\begin{align}
   &\left( \partial_x-\partial_y\right)\g_2^{\{1\}}(x,y)\Big|_{x=y^-}  
=\nonumber\\
&2\partial_x \g_1^{\{0\}}(x)\partial_x \g_1^{\{1\}}(x)
          -2\varepsilon  \partial_x \g_1^{\{0\}}(x)\g_1^{\{0\}}(x)(\g_1^{\{0\}}(x)+\varepsilon).
\end{align}
Thus, assuming that $\mathfrak{g}_{2}^{\{1\}}$ is a polynomial in $x$ and $y$, it can be at most of fourth order in each variable. Imposing the boundary conditions at $x=0$ and $y=1$, we end up with the ansatz
\begin{equation}
\mathfrak{g}_{2}^{\{1\}}(x,y)=x(1-y)\left(a_0 + a_1 x + a_2 y + b_1 x^2 + b_2 y^2 + b_3 xy \right),
\end{equation}
which leaves six unknown coefficients. The remaining bulk equations, together with the boundary condition at $x=y-0^+$, form an overdetermined system. Remarkably, all equations are mutually consistent, allowing the solution to be determined explicitly as
\begin{align}\label{eq:g2-u1}
\mathfrak{g}_{2}^{\{1\}}(x,y)=\left(\Delta \bar n\right)^{2}x\left(1-y\right)&\Big[x+y-2+\frac{2\sign\Delta\bar n}{3}\nn
&\times\left(2+y-2\left(x^{2}+y^{2}\right)\right)\Big],
\end{align}
valid for $x<y$ and where $\Delta \bar n=\bar n_0 - \bar n_N$. Eq.~\eqref{eq:g2-u1} enables the computation of the third cumulant generated by $\delta F_{\rm qu}(u)$, given by
\be
\frac1N\frac{\de^3}{\de u^3} \delta F_{\rm qu}(u)\Big\vert_{u=0} =\frac{(\Delta \bar n)^2}{N^2}  \big(1 -\frac{13\sign\Delta \bar n}{15}\big).
\ee
{This result provides an explicit and quantitative expression for the quantum correction to the third-order cumulant, which is suitable for experimental or numerical verification, as discussed in the introduction for the current variance. We further expect this contribution to coincide, up to an overall minus sign, with the $1/N$ bulk finite-size corrections obtained in the classical process, see discussion in Sec.~\ref{sec:exact-calc}. A systematic expansion to higher orders in small $u$, or analytical or numerical solutions of Eq.~\eqref{eq:secondorder_g}, would enable the reconstruction of the full statistics of the quantum corrections.
}
\section{Conclusion}
\label{sec:conclusion}

In this work we investigated the statistics of integrated currents in the QSSEP/QSSIP models, combining the two-time measurement protocol with a biased quantum evolution. Our main result is the identification of a clear separation between the leading contribution to the cumulant-generating function and its subleading corrections. To leading order in the system size, the current large deviation function coincides with that of the associated noise-averaged (classical) process. This establishes, in a precise sense, a form of {\it classical typicality} for current fluctuations in these noisy quantum systems: although the dynamics is genuinely quantum at the microscopic level, the dominant contribution to transport fluctuations is fully captured by an effective classical description. {Our derivation relies on the Gaussian structure of the QSSEP/QSSIP dynamics and therefore does not directly extend to generic quantum diffusive systems. Nonetheless, we expect this notion of classical typicality of charge~\cite{Hruza2023,Bernard2025} and current fluctuations, at leading order in the system size, to hold in generic diffusive quantum systems. 

This expectation is motivated by the hydrodynamic character of the leading contribution and by related evidence from charge-conserving random circuits~\cite{FCSMFTDeNardis}. In such systems, one may expect diffusion and local conservation laws to control the leading large-scale current statistics, while quantum coherence contributes at subleading scales. This extension remains conjectural and has yet to be proven beyond the QSSEP/QSSIP models considered here.}\\

{We also emphasize that the classical form of the noise-averaged dynamics alone does not imply $\E[\log Z_{t,w}(u)]\sim\log\E[Z_{t,w}(u)]$. Our exact microscopic calculation establishes this relation to leading order in the system size. It also demonstrates agreement with the effective MFT description and provides, to the best of our knowledge, the first derivation of the MFT current statistics directly from the microscopic noisy quantum dynamics. The self-averaging analysis further establishes convergence to this classical behavior for a typical realization of the noise.}\\

Beyond this leading behavior, we have shown that quantum effects enter at subleading order. By carefully tracking the It\=o corrections arising from the stochastic evolution, we isolated a quantum contribution to the cumulant-generating function, which is absent at the classical level and scales as ${\cal O}(N^{-1})$.  We argued that this term is distinct from the finite-size corrections already present in the classical process~\cite{Derrida2004}, and that the two contributions combine to generate the fluctuation statistics of currents in the quantum system. {Remarkably, the presence of this quantum term leads to an exact cancellation of the contributions that would generate bulk finite-size corrections in the current cumulants.} Our results place the QSSEP/QSSIP models in a broader hydrodynamic context. On the one hand, the leading-order agreement with MFT confirms that these systems fall within the standard framework of diffusive transport, despite their quantum nature. On the other hand, the explicit form of the quantum correction highlights the limits of a purely classical description and quantifies how quantum features can survive in transport properties at large (but finite) scales.\\

{We also emphasize that our predictions for the quantum corrections can be experimentally or numerically investigated using a well-defined protocol, provided sufficient control over the noise realizations in the model can be achieved. In this perspective, we highlight our result for the quantum correction to the current variance in eq.~\eqref{eq:res-ucf}, which is particularly suited for a direct and quantitative verification.\\

A closely related and natural question is whether the structure of the quantum corrections identified here is specific to the noninteracting nature of the QSSEP/QSSIP models. While this possibility cannot be excluded a priori, several indications suggest a broader validity. Random circuit models provide examples of genuinely interacting systems displaying diffusive transport, while related constructions, such as for noisy XXZ-type models~\cite{Bauer2017}, point toward similar mechanisms at work beyond free theories. Finally, it would be interesting to explore whether analogous quantum corrections arise in ballistic or weakly diffusive regimes, and to what extent the present approach can be interfaced with generalized hydrodynamics or fluctuating hydrodynamic theories in those settings~\cite{Doyon2018,Doyon2019,Doyon2020,Doyon2023,Doyon2023b}. In this broader perspective, the development of a quantum extension of MFT~\cite{Bernard2021}, possibly formulated within a variational framework, appears as a natural and promising direction.}\\

We expect that the framework developed here provides a useful starting point for addressing all these questions, and more generally for clarifying the roles of classical and quantum contributions to nonequilibrium fluctuations in extended open quantum systems.
\\[.1cm]

\textit{Acknowledgements.}~---~We acknowledge J. Costa, A. De Luca, F. Carollo, and L. Hruza, for useful discussions.  This work was supported by the CNRS, the ENS, the ANR project ESQuisses under contract number ANR-20-CE47-0014-01, and by the Simons Collaboration ``Probabilistic Paths to QFT''. SS acknowledges support from the MSCA Grant No. 101103348 (GENESYS). This work has been partially funded by the European Union. Views and opinions expressed are however those of the author(s) only and do not necessarily reflect those of the European Union or the European Commission. Neither the European Union nor the European Commission can be held responsible for them.
 
\appendix

\section{Two-time measurement scheme}\label{app:two-time-measurem}
In this section we present a microscopic derivation of
eqs.~\eqref{eq:bias-bath} and \eqref{eq:dH-bias} appearing in the main
text. We begin by considering the total Hamiltonian of a generic system
coupled to a reservoir,
\begin{equation}
\hat H=\hat H_{S}+\hat H_{R}+\hat H_{SR},
\end{equation}
where $\hat H_{S/R}$ denotes the system/reservoir Hamiltonian, while the
additional term describes their mutual interaction. For simplicity, we
consider a free bosonic reservoir in one spatial dimension with periodic
boundary conditions and total length $L_{R}=N_{R} a$, where $N_{R}$ is the
number of sites in the bath and $a$ the lattice spacing. The associated
Fourier modes are indexed by $k=\frac{2\pi}{N_{R}}m$, $m\in\mathbb{Z}$,
and the dispersion relation is denoted by $\omega\left(k\right)$. The
reservoir Hamiltonian can then be written as
\begin{equation}
\hat H_{R}:=\sum_{k}\omega\left(k\right)\tilde{n}_{k},
\end{equation}
where the tilde indicates Fourier-space operators and $\tilde n_{k}$ is
the occupation number operator of mode $k$.

We assume that the system is coupled to the environment, in position
space, through
\begin{equation}
\hat H_{SR}:=J\left(\hat {S}^{\dagger}\ \hat A_{0}+\text{h.c.}\right),
\end{equation}
where $\hat {S}^{\dagger}$ is an operator acting on the system, left
unspecified at this stage, and $\hat A_{0}$ acts on the bath. Physically,
this term describes the creation of a boson in the reservoir when $\hat
S$ acts on the system, and viceversa for $\hat S^\dagger$. The operator $\hat A_{0}$ admits
the Fourier decomposition
\begin{equation}
\hat A_{0}=\frac{1}{\sqrt{N_{R}}}\sum_{k}\tilde{A}_{k},
\end{equation}
where $N_{R}$ is the number of bath modes. \\

The quantity of our interest is the total charge {\it in the reservoir},
\begin{equation}
\hat{Q}:=\sum_{j=1}^{N_R}\hat{n}_{j}.
\end{equation}
We define the integrated current $Q(t)$ as the difference between the
measured value of $\hat{Q}$ at time $t$, denoted $q(t)$, and its value at
time $0$, $q_{0}$. For simplicity, we assume that at $t=0$ the value
$q_{0}$ is obtained with unit probability (for instance, by starting
from an empty reservoir). \\

Let $\hat\Pi_{q}=|q\rangle\langle q|$ denote the projector onto the sector
with fixed charge $q$. The probability $\text{Prob}\left[Q(t)\right]$ is
then given by
\begin{equation}
\text{Prob}\left[Q(t)\right]={\rm tr}\left(\hat\Pi_{q_{t}}\ \hat U_{t}\ \hat\Pi_{q_{0}}\ \hat\rho_{0}\ \hat\Pi_{q_{0}}\ \hat U_{t}^{\dagger}\ \hat\Pi_{q_{t}}\right),
\end{equation}
with evolution operator
\[
\hat U_t=\exp\left[-it\left(\hat H_{S}+\hat H_{R}+\hat H_{SR}\right)\right].
\]

The corresponding moment-generating function is obtained by summing over
all possible outcomes,
\begin{align}
 Z_{t}\left(\zeta\right):&=\sum_{Q_{t}}\ \text{Prob}\left[Q(t)\right] \ e^{-\zeta Q_{t}}\\
 & =\sum_{Q_{t}}{\rm tr}\left(\hat\Pi_{q_{t}}\ \hat U_{t}\ \hat\Pi_{q_{0}}\ \hat\rho_{0}\ \hat\Pi_{q_{0}}\ \hat U_{t}^{\dagger}\ \hat\Pi_{q_{t}}\right)e^{-\zeta\left(q_{t}-q_{0}\right)}\nonumber \\
 & ={\rm tr}\left[\hat\rho_t(\zeta)\right],\nonumber
\end{align}
where
\begin{equation}
\hat\rho_t(\zeta):= \hat U_{t,\zeta}\ \widetilde{\rho}_{0} \ \hat U_{t,-\zeta}^{\dagger}
\ee
and $\widetilde{\rho}_{0}:=\hat\Pi_{q_{0}}\ \hat\rho_{0}\ \hat\Pi_{q_{0}}$,  equal to $\hat\rho_0$ if one choose an initial state with defined charge, $[\hat\rho_0,\hat Q]=0$. We have also introduced the biased evolution operator
\begin{equation}
U_{t,\zeta}:=e^{\frac{\zeta\hat{Q}}{2}}\ \hat U_{t} \ e^{-\frac{\zeta \hat{Q}}{2}}
=  \exp\left[-it\left(\hat H_{S}+\hat H_{R}+\hat H_{SR,\zeta}\right)\right],
\end{equation}
where
\begin{equation}
\hat H_{{\rm SR},\zeta}=J\left(e^{-\frac{\zeta}{2}} \hat {S}^{\dagger}\hat{A}_{0}
+e^{\frac{\zeta}{2}} \hat A_0^\dagger \hat {S} \right)
\end{equation}
following from  $\left[\hat{Q},\hat{A}_{0}\right]=-\hat{A}_{0}$. This reproduces eq.~\eqref{eq:bias-bath}. From this result, we can now present a standard microscopic derivation of the stochastic hopping terms, see also Refs.~\cite{HudsonParthasarathy_QuantumIto,BreuerPetruccione_book,GardinerZoller_quantumnoise}. To simplify the discussion, we take the continuum limit for the energy
modes of the bath. Introducing
\be
\tilde{A}_{m}:=\sqrt{\frac{2\pi}{L_{B}}}\,\hat A_{\omega},
\qquad
\left[\hat A_{\omega},\hat A_{\omega'}^{\dagger}\right]
\underset{L_{R}\to\infty}{=}\delta\left(\omega-\omega'\right),
\ee
the reservoir Hamiltonian takes the form
\begin{equation}
\hat H_{R}:=\int\frac{d\omega}{\omega'\left(k\right)}\,
\omega\; \hat A_{\omega}^{\dagger}\hat A_{\omega},
\end{equation}
while the local bath operator reads
\begin{align}
\hat A_{0} & =\sqrt{\frac{a}{2\pi}}\int\frac{d\omega}{\omega'\left(k\right)} \hat A_{\omega}.
\end{align}

We now assume a linear dispersion relation for the bath,
\begin{equation}
\omega=ck,
\end{equation}
and set $c=1$ for simplicity. Under this assumption the reservoir Hamiltonian
reduces to
\begin{equation}
\hat H_{R}=\int d\omega\,\omega\; \hat A_{\omega}^{\dagger}\hat A_{\omega},
\end{equation}
and the system-reservoir coupling becomes
\begin{align}
H_{SR,\zeta}
& :=\sqrt{\gamma}
\int\frac{d\omega}{\sqrt{2\pi}}
\left(e^{-\frac{\zeta}{2}} \hat {S}^{\dagger}\hat{A}_{\omega}
+e^{\frac{\zeta}{2}} \hat A_\omega^\dagger \hat {S} \right),
\end{align}
with $\gamma:= aJ^2$. We next move to the interaction picture with respect to the bath,
\begin{equation}
\hat\rho_{t}^{{\rm I}}(\zeta):=
e^{itH_{R}}\hat\rho_{t}(\zeta)e^{-itH_{R}}.
\end{equation}
In this representation the evolution of the density matrix reads
\begin{align}
 d\hat\rho_{t}^{{\rm I}}(\zeta)
 &=-i\left[\hat H_{S},\hat\rho_{t}^{{\rm I}}(\zeta)\right]
 -i\hat H_{SR,\zeta}^{{\rm I}}(t)\hat\rho_{t}^{{\rm I}}(\zeta)
 +i\hat\rho_{t}^{{\rm I}}(\zeta)\hat H^I_{SR,-\zeta}(t),
\end{align}
where
\begin{equation}
\hat H_{SR,\zeta}^{{\rm I}}(t):=
e^{itH_{R}}\ H_{{ SR},\zeta}\ e^{-itH_{R}}.
\end{equation}
The reservoir operators evolve as
\begin{align}
\hat A_{\omega}\left(t\right) & =e^{i\omega t}\hat A_{\omega},\\[4pt]
\hat A_{\omega}^{\dagger}\left(t\right) & =e^{-i\omega t}\hat A_{\omega}^{\dagger}.
\end{align}

A key observation is that
\begin{align}
\left[\int\frac{d\omega}{\sqrt{2\pi}}\hat A_{\omega}\left(t\right),
\int\frac{d\omega'}{\sqrt{2\pi}}\hat A_{\omega'}^{\dagger}\left(t'\right)\right]
& =\delta\left(t-t'\right),
\end{align}
which shows that $\int\frac{d\omega}{\sqrt{2\pi}}\hat A_{\omega}\left(t\right)$ behaves as a
white noise. This motivates rewriting the Hamiltonian increment
as
\begin{equation}
dH_{{ SR},\zeta}\left(t\right)
=\sqrt{\gamma}\left(e^{-\frac{\zeta}{2}} \hat{S}^{\dagger}d\hat{w}(t) + e^{\frac{\zeta}{2}} d\hat{w}^\dagger(t) \hat {S} \right),
\label{eq:dH_SB_lambd}
\end{equation}
with
\begin{equation}
d\hat{w}(t):=dt\int\frac{d\omega}{\sqrt{2\pi}}\hat A_{\omega}\left(t\right).
\end{equation}

The time-step increment of the biased density matrix is then expressed as
\begin{equation}
\hat\rho_{t+dt}(\zeta)
=e^{-i\left(\hat H_{s} dt+ d\hat H_{{ SR},\zeta}\right)}\,
\hat\rho_{t}(\zeta)\,
e^{i\left(\hat H_{S}dt+d\hat H_{{SR},-\zeta}\right)}.
\end{equation}

The next step is to use the reservoir degrees of freedom to define a statistical
average for the operator $d\hat{w}(t)$. We assume that the coupling to the
system can be neglected when evaluating reservoir correlations, so that the
total density matrix factorizes as
\begin{equation}
\hat\rho\left(\zeta\right)
=\hat\rho_{S}(\zeta) \otimes\hat\rho_{R}.
\end{equation}
If $\hat\rho_{R}$ is diagonal in the energy eigenbasis, it remains stationary
in the interaction picture. The bath state then fixes the statistics of
the stochastic increment $d\hat{w}(t)$ as
\begin{align}
\mathbb{E}\left[d\hat{w}(t)\right]
& :={\rm tr}\left(\hat\rho_{R}\ d\hat{w}(t)\right)=0,\\[4pt]
\mathbb{E}\left[d\hat w(t)d\hat{w}^\dagger(t')\right]
& :={\rm tr}\left(d\hat{w}(t) \hat\rho_{R} \ d\hat{w}^{\dagger}(t') \right)\nonumber\\
&:=\begin{cases}
\alpha\, dt & \text{if }t=t',\\
0 & \text{otherwise},
\end{cases}\nonumber \\[4pt]
\mathbb{E}\left[d\hat{w}(t')^\dagger d\hat w(t)\right]
& :={\rm tr}\left(d\hat{w}^{\dagger}(t')\hat\rho_{R}\ d\hat w(t)\right)\nonumber\\
&:=\begin{cases}
\beta\, dt & \text{if }t=t',\\
0 & \text{otherwise}.
\end{cases}\nonumber
\end{align}
Note that, when $\alpha\neq\beta$, care must be taken with the relative ordering of
$d\hat{w}$, $d\hat w^{\dagger}$, and $\hat\rho_{R}$. These relations can finally be promoted to It\=o rules in the standard
way. The key step is that products of pairs of $d\hat{w}(t)$ and
$d\hat{w}^{\dagger}(t)$ can be replaced by their expectation values. The
resulting stochastic process differs microscopically from the original one but converges to as $dt\to0$ \cite{Oksendal2003}. At this point it
is also convenient to forget the operator nature of the noise and to identify the operators $d\hat{w}(t)$, $d\hat{w}^{\dagger}(t)$ with the $\mathbb{C}$ numbers $dw(t)$, $d\overline{w}(t)$.\\

To recover the biased dynamics of QSSEP/QSSIP in eq.~\eqref{eq:dH-bias}, one may proceed in
two ways. After setting $\gamma=1$, one may choose $\hat {S}$ to be a boundary operator, setting
\begin{align}
\hat {S} & =\hat c_{0/N},\\
\left(\alpha,\beta\right) & =\left(\alpha_{0/N},\beta_{0/N}\right),
\end{align}
and systematically average over the bath degrees of freedom. Alternatively, one may choose $\hat S$ as a link operator,
\begin{align}
\hat {S} & =\hat\ell_j,\\
\alpha & =\beta=1,
\end{align}
so that a boson is created in the reservoirs whenever a particle hops from site
$j$ to site $j+1$ (cf Fig.~\ref{fig:setup}).\\

As a side remark, the unbiased case ($\zeta=0$) in the above
construction provides a microscopic derivation of the QSSEP/QSSIP models, see also Ref.~\cite{Bauer2017}.

\section{Gaussian-preserving biased dynamics}
\label{app:gaussian-prop}
In this Appendix, we show that the biased quantum dynamics~\eqref{eq:dyn-bias} preserves the Gaussian character of an initially Gaussian state. We start by introducing the following notation for generic Gaussian states
\begin{equation}
\hat\rho[K] := \frac{\hat\sigma[K]}{Z[K]}, \quad \hat\sigma[K]:= \exp\left(\sum_{i,j=0}^N \hat{c}_i^\dagger K_{ij} \hat{c}_j\right),
\end{equation}
with partition function $Z[K]$ and  two-point function $G[K]$ given by
\be\label{eq:Z-and-G-gauss}
Z[K] := \frac{1}{\big(\det[1-\sign e^{K}]\big)^\sign}, \quad G[K] := \frac{e^K}{1-\sign e^K}.
\ee
Following Refs.~\cite{two_times_Esposito_2009,Landi2024current}, one can write the biased density matrix resulting from a two-time measurement protocol as
\be
\hat\rho_{t,w}(u)= e^{\frac{u}{2} \hat A_\mu} \, \Phi_{t,w}\left[  e^{-\frac{u}{2} \hat A_\mu} \, \hat\rho_0 \,  e^{-\frac{u}{2} \hat A_\mu} \right]  e^{\frac{u}{2} \hat A_\mu},
\ee
where $\Phi_{t,w}[ \ ]$ is the dynamical map generating the biased dynamics \eqref{eq:dyn-bias}; $\hat{A}_\mu := \sum_j \mu_j \hat n_j$ and $\mu_j :=\mu_0 +\sum_{i=0}^{j-1} \lambda_i$ up to a constant $\mu_0$ that does not affect the evolution in \eqref{eq:dyn-bias}. \\

The state is initially prepared in a Gaussian state $\hat\rho_0=\hat\rho[K_0]$. Performing the first projective measurement at $t=0$ 
\begin{equation}
\hat\sigma[K(u;0)] := e^{-\frac{u}{2} \hat A_\mu} \, \hat\sigma[K_0] \, e^{-\frac{u}{2} \hat A_\mu}
\end{equation}
where
\begin{equation}\label{eq:composition-polished}
K(u;0) := \log\Big[ e^{-\frac{u}{2} \vec\mu } \, e^{K_0}\, e^{-\frac{u}{2} \vec\mu} \Big], 
\quad \vec\mu := \mathrm{diag}(\mu_0,\dots,\mu_N).
\end{equation}
This follows from the observation that the map
\begin{equation}
X \mapsto \mathcal{O}_X := \vec{c}^{\,\dagger} X \, \vec{c}
\end{equation}
forms a representation of $\mathfrak{gl}(N+1)$. Consequently, operator exponentials satisfy the composition law
\begin{equation}
e^{\mathcal{O}_X} \, e^{\mathcal{O}_Y} \, e^{\mathcal{O}_X} = e^{\mathcal{O}_{\log(e^X e^Y e^X)}},
\end{equation}
which leads directly to~\eqref{eq:composition-polished}. The stochastic time evolution generated the Gaussian-preserving dynamical map $\Phi_{t,w}[\ ]$ then gives
\begin{equation}
\hat\sigma[K^{(w)}(u;t)] := \Phi_{t,w}[\hat\sigma[K(u;0)]].
\end{equation}
Applying the second projective measurement at time $t>0$ yields
\begin{equation}
\hat\sigma[\tilde{K}^{(w)}(u;t)] := e^{\frac{u}{2}\hat{A}_\mu}\, \hat\sigma[K^{(w)}(u;t)] \, e^{\frac{u}{2}\hat{A}_\mu} 
\end{equation}
with
\begin{equation}
\tilde{K}^{(w)}(u;t) := \log\Big[ e^{\frac{u}{2} \vec\mu} \, e^{K^{(w)}(u;t)} \, e^{\frac{u}{2} \vec\mu} \Big].
\end{equation}
Hence, the biased density matrix
\be
\hat\rho_{t,w}(u) =\frac{\hat\sigma[\tilde{K}^{(w)}(u;t)]}{Z[K_0]}
\ee
 remains a Gaussian state at all times, and Wick’s theorem applies. Finally, we note that the moment-generating function $Z_{t,w}(u)=\tr\big(\hat\rho_{t,w}(u)\big)$ can be expressed in determinant form as
\begin{equation}
Z_{t,w}(u) = \frac{Z[\tilde{K}^{(w)}(u;t)]}{Z[K_0]} 
= \left( \frac{\det\left[1+\sign G^{(w)}(u;t)\right]}{\det\left[1+\sign G_0\right]} \right)^\sign,
\end{equation}
where $G^{(w)}(u;t)= G[\tilde{K}^{(w)}(u;t)]$ and $G_0=G[K_0]$. Note that the argued self-averaging condition of current fluctuations implies that the trace of the random matrix
\be
\Omega_{t,w}(u):= \sign\log \left[ \frac{1+\sign G^{(w)}(u;t) }{1+\sign G_0} \right]
\ee
is self-averaging.

\section{Convergence of the averages}\label{app:convergence}
In this Appendix, we provide a derivation of eq.~\eqref{eq:rel-avg}, relating the quantum $\E[\langle \ \rangle_{w,u}]$ and the classical  $\langle \ \rangle_{{\rm cl},u}$ averages for a generic product of density operators. We first recall that
\begin{align}
\frac{1}{dt}\E\left[\frac{dZ_{t,w}}{Z_{t,w}}\right](u)
=\sum_{j=0}^{N-1} \E\Big[&
\langle \hat{n}_{j+1}(1+\sign \hat n_j)(e^{-u\lambda_j}-1)\nn
&+\hat{n}_{j}(1+\sign \hat n_{j+1})(e^{u\lambda_j}-1)\rangle_{w,u}\Big],
\end{align}
and, denoting $Z^{\rm cl}_t(u):=\E[Z_{t,w}(u)]$,
\begin{align}
\frac{1}{dt}\frac{dZ^{\rm cl}_{t}}{Z^{\rm cl}_{t}}(u)
=\sum_{j=0}^{N-1} &\langle \hat{n}_{j+1}(1+\sign \hat n_j)(e^{-u\lambda_j}-1)\nn
&+\hat{n}_{j}(1+\sign \hat n_{j+1})(e^{u\lambda_j}-1)\rangle_{{\rm cl},u}.
\end{align}
Both quantities scale as ${\cal O}(1/N)$ {for $\lambda_j\sim {\cal O}(1/N)$ (cf eq.~\eqref{eq:gauge})}. \\

We now consider a generic operator $\hat O=\hat n_{i_1}\dots \hat n_{i_k}$ made of a string of density operators, and recall the definitions of the
two averages,
\begin{align}
\langle \hat O\rangle_{{\rm cl},u}:=\frac{\mathbb{E}\!\left[{\rm tr}\!\left(\hat\rho_{t,w}(u) \hat O\right)\right]}{Z^{\rm cl}_t(u)}, \quad
\langle \hat O\rangle_{w,u}:=\frac{{\rm tr}\!\left(\hat\rho_{t,w}(u) \hat O\right)}{Z_{t,w}(u)}.
\end{align}
The equations of motion in the two cases
can then be written as follows. For the noise-averaged process, using the standard Leibniz rule, one has
\be
d\langle \hat O\rangle_{{\rm cl},u}
=dt \langle {\cal L}_u^*(\hat O)\rangle_{{\rm cl},u}
-\langle \hat O\rangle_{{\rm cl},u}\frac{dZ^{\rm cl}}{Z^{\rm cl}},
\ee
and, imposing stationarity in the steady state,
\be\label{eq:stat-cl}
\langle \hat O\rangle_{{\rm cl},u}
=\frac{dt \langle {\cal L}_u^*(\hat O)\rangle_{{\rm cl},u}}{dZ^{\rm cl}/Z^{\rm cl}}.
\ee

On the other hand, for the stochastic process, using It\=o calculus, one has
\be
d\E[\langle \hat O\rangle_{w,u}]
=\E\Big[dt\langle {\cal L}_u^*(\hat O)\rangle_{w,u}
-\langle \hat O\rangle_{w,u}\frac{dZ_{w}}{Z_{w}} + \Delta O \Big],
\ee
with It\=o correction
\be
\Delta O:= -\langle d{\cal H}^*_{w,u}(\hat O)\rangle_{w,u}\frac{dZ_{w}}{Z_{w}}+\langle \hat O\rangle_{w,u}\left(\frac{dZ_{w}}{Z_w}\right)^2.
\ee
In the stationary limit, this allows us to isolate the steady-state expectation
\begin{align}
\E\!\left[\langle \hat O\rangle_{w,u}\right]
=\mathbb{E}\!\left[\frac{dZ_{w}}{Z_{w}}\right]^{-1}
\mathbb{E}\left[dt\langle {\cal L}_u^*(\hat O)\rangle_{w,u} + \Delta O\right]\big(1+{\cal O}(N^{-2})\big),
\label{eq:av_expressions}
\end{align}
where we used  the factorization property (cf.~axioms \hyperlink{axioms}{\it(iii-iv)}), 
\be
\E\left[\langle \hat O\rangle_{w,u} \frac{dZ_w}{Z_w}\right]=\E\left[\langle \hat O\rangle_{w,u}\right]\E\left[ \frac{dZ_w}{Z_w}\right]\big(1+{\cal O}(N^{-2})\big).
\ee

The difference between the classical \eqref{eq:stat-cl} and quantum \eqref{eq:av_expressions} steady-state expectation values is thus
contained in the quantity $\E[\Delta O]/\E[dZ_w/Z_w]$. \\

A direct inspection of this quantity, using the explicit form of the biased
Hamiltonian together with Wick’s theorem, shows that
\be
\frac{1}{\E[\langle\hat O \rangle_{w,u}]} \ \frac{\E[\Delta O]}{\E[dZ_w/Z_w]}={\cal O}(N^{-2})
\ee
since the leading contributions for the two terms in $\Delta O$ cancel for any operator $\hat O$ given by a product of
densities. As a consequence, eq.~\eqref{eq:av_expressions} reduces to
\be \label{eq:stat-q}
\mathbb{E}\!\left[\langle \hat O\rangle_{w,u}\right]
=
\frac{\E\left[dt\langle {\cal L}_u^*(\hat O)\rangle_{w,u}\right]}{\E[dZ_w/Z_w]}\big(1 +{\cal O}(N^{-2})\big).
\ee
By comparing \eqref{eq:stat-cl} and \eqref{eq:stat-q}, property~\eqref{eq:rel-avg} of the main text then follows immediately, since the expressions for the quantum and classical averages coincide to order ${\cal O}(N^{-2})$, up to the replacement of the corresponding expectation values. \\

We thus conclude that the steady-state average $\E[\langle\;\rangle_{w,u}]$ converges in law to $\langle\;\rangle_{{\rm cl},u}$ at leading order in the system size.  This provides a complementary, yet equivalent to the direct calculation presented in the main text, justification of the classical typicality observed in these systems in the limit $N\to\infty$.
\section{Exact calculation of low-order cumulants}\label{app:low-order}
In this Appendix, we present an exact microscopic derivation of the lowest-order cumulants of $\E[F_{{\rm qu},w}(u)]$ in \eqref{eq:Ftmp2}. We shall show that the final results are independent of the specific choice of the weight function $\lambda_j$. In what follows, we find it convenient to consider the generating function $\E[F_{{\rm qu},w}(u)]/N$, i.e. with an overall $1/N$ factor included so as to match the conventions used in Refs.~\cite{Derrida2004,Derrida2007} for the SSEP.\\

We begin by considering the bias acting on a single link $k$ connecting sites $k\to k+1$. This case is obtained from \eqref{eq:Ftmp2} setting $\lambda_j =\delta_{jk}$, and leads to the generating function
\be\label{eq:F-single-probe}\begin{split}
\frac{\E[F_{{\rm qu},w}(u)]}{N}=& (e^{-u}-1)\left( n_{k+1}+ \sign m_{k,k+1}\right)\nn
&+(e^{u}-1)\left( n_{k}+ \sign m_{k+1,k}\right)\\
&+ 2\sign\left(\cosh(u)-1\right) n_kn_{k+1},
\end{split}\ee
with
$m_{i,j}(u):=\E[G_{i,i}^{(w)}G_{j,j}^{(w)}]^c(u)$ and $n_j(u):=\E[n_j^{(w)}](u)$. Expanding for small $u$ yields
\begin{align}\label{F-u-lowest}
\frac{\E[F_{{\rm qu},w}(u)]}{N}= & \,u\big( n_{k}-n_{k+1}\big)\nn
&+\frac{u^2}{2}\Big[n_k+n_{k+1}+2\sign \big( m_{k,k+1}+ n_k n_{k+1}\big)\Big]\nn
&\quad  +\frac{u^3}{6}\big(n_{k}-n_{k+1}\big)+{\cal O}(u^4).
\end{align}
This expansion for $F(u)$ must be complemented with
\begin{align}\label{eq:exp-corr}
&n_j=n_j^{\{0\}} + u \, n_j^{\{1\}}  + u^2 n_j^{\{2\}} + {\cal O}(u^3),\\
&m_{i,j}=m_{i,j}^{\{0\}} + u \, m_{i,j}^{\{1\}}  + {\cal O}(u^2),
\end{align}
where $n_{j}^{\{q\}}$ and $m_{i,j}^{\{q\}}$ denote bias corrections to cumulants at order ${\cal O}(u^q)$. For $q>0$, these functions are to be determined, while the zeroth-order terms (corresponding to the unbiased case) are \cite{Bernard2019,Derrida2007,Bernard2025}
\be\label{eq:unbias-n}
n_{j}^{\{0\}} = \frac{\bar n_0(N+b-j)+\bar n_N(j+a)}{a+b+N},
\ee
and
\be\label{eq:unbias-m}\begin{split}
m_{i,j}^{\{0\}}=-\frac{(\bar n_N-\bar n_0)^2(i+a)(N-j+b)}{(N+a+b-1)(N+a+b)^2(N+a+b+1)}
\end{split}\ee
with
\be\label{eq:def-a,b}
a:=\big(\beta_0-\sign \alpha_0\big)^{-1},\quad
b:=\big(\beta_N-\sign \alpha_N\big)^{-1}.
\ee
The first-order current cumulant follows directly from the unbiased correlations. One finds, independently of the biased link $k$,
\be
\frac{\E[F_{{\rm qu},w}(u)]}{N}=u \cum_1+{\cal O}(u^2) 
\ee
with
\be\label{eq:kappa1}
\cum_1:=\,\frac{\bar n_0-\bar n_N}{N+a+b} 
\;\;\overset{N\to\infty}{\simeq}\;\; \,\frac{\bar n_0-\bar n_N}{N}.
\ee
This shows that the average number of transferred particles is determined by the imbalance of the boundary densities.\\

To second order in $u$, one obtains
\be
\frac{\E[F_{{\rm qu},w}(u)]}{N}-u\cum_1 =u\big(n_k^{\{1\}}-n_{k+1}^{\{1\}}\big)+\dfrac{u^2}{2}\,T_{k,k+1},
\ee
with $T_{i,j} := n_i+n_j+2\sign \big(m_{i,j}+n_i n_j\big)$. Summing over the links and rescaling $u\to u/N$, one finds~\cite{Derrida2004}
\begin{align}\label{eq:variance-tmp}
\frac{\E[F_{{\rm qu},w}(u)]}{N}-u\cum_1
= \frac{u^2}{2N^2} \sum_{k=0}^{N-1}T_{k,k+1},
\end{align}
 where the dependence on $n^{\{1\}}_k$ cancels since $n_{0/N}^{\{q>0\}}={\cal O}(N^{-1})$. Eq.~\eqref{eq:variance-tmp} can now be evaluated using \eqref{eq:unbias-n}–\eqref{eq:unbias-m}. In the large-$N$ limit, this yields
\be
\frac{\E[F_{{\rm qu},w}(u)]}{N} =  u\cum_1 + \frac{u^2}{2}\cum_2 + {\cal O}(u^3),
\ee
with current variance
\be\label{eq:kappa2}
\cum_2 =\frac{1}{N}\left[\bar n_0+\bar n_N+\frac{2\sign}{3}\Big(\bar n_0^2+\bar n_0 \bar n_N+\bar n_N^2\Big)\right] +{\cal O}(N^{-2}).
\ee

By expanding \eqref{F-u-lowest}--\eqref{eq:exp-corr} to order ${\cal O}(u^3)$ gives
\be\begin{split}
\frac{\E[F_{{\rm qu},w}(u)]}{N}-&\sum_{q=1}^2\frac{u^q\cum_q}{q!}=\\ &u\big(n_k^{\{2\}}-n_{k+1}^{\{2\}}\big)+\frac{u^2}{2}T^{\{1\}}_{k,k+1}+\frac{u^3}{6}\big(n^{\{0\}}_k-n^{\{0\}}_{k+1}\big)
\end{split}
\ee
with $T_{i,j}^{\{1\}}:=n^{\{1\}}_i + n^{\{1\}}_j+2\sign \big( m^{\{1\}}_{i,j}+ n^{\{1\}}_i n^{\{0\}}_j+n^{\{0\}}_i n^{\{1\}}_j\big)$. However, summing over $k$ and rescaling the bias amplitude $u\to u/N$ removes the dependence on $n_j^{\{2\}}$ but leaves the contribution from $m_{i,j}^{\{1\}}$. In general, the computation of the $q$-th current cumulant needs to know $n_j$ up to ${\cal O}(u^{q-2})$ and $m_{i,j}$ up to ${\cal O}(u^{q-3})$, and in turn all correlations up to the $(q-1)$-point density correlation, $\E[\tr(\hat\rho_{t,w}\hat{n}_{i_1}\dots \hat{n}_{i_{q}})]^c$, to order zero~\cite{Derrida2007}. This calls for a systematic order-by-order expansion in small $u$.  \\

For large $N$, the analysis simplifies since (cf.~axiom \hyperlink{axioms}{\it(iv)}) 
\be
m_{i,j} \sim {\cal O}(N^{-2}).
\ee
Using this, and after summing over $k$ and rescaling $u\mapsto u/N$, eq.~\eqref{eq:F-single-probe}, to leading order in $1/N$, becomes
\begin{align}\label{eq:F-to-expand}
\frac{\E[F_{{\rm qu},w}(u)]}{N} &=\sum_q \frac{u^q \cum_q}{q!} \nn
&\overset{N\to\infty}{\simeq}\frac{-u}{N} \g_1(x)\Big\vert_{x=0}^{x=1} +\frac{u^2}{N}\int_0^1 dx \  \g_1(x)\left(1+\sign \g_1(x)\right).
\end{align}
Note that this expression is equivalent to that obtained choosing uniform weights, i.e., $\xi(x)\equiv 1$, in eq.~\eqref{eq:F-general}. Indeed, expanding the function $\g_1(x)$ as
\be\label{eq:n-expansion-u} 
\g_1(x)=\g_1^{\{0\}}(x)+ u\g_1^{\{1\}}(x) + u^2 \g_1^{\{2\}}(x) + {\cal O}(u^3),
\ee
with zeroth-order profile
\be\label{eq:g0}
\g_1^{\{0\}}(x)=\bar n_0(1-x)+x \bar n_N,
\ee
given by the unbiased solution \cite{Bernard2019}, one can then treat the problem systematically via a small-$u$ expansion. The first two cumulants in eqs.~\eqref{eq:kappa1} and \eqref{eq:kappa2} follow directly from \eqref{eq:F-to-expand} upon inserting the unbiased function \eqref{eq:g0}. Higher-order cumulants, instead, require the systematic evaluation of the bias-induced corrections to $\mathfrak{g}_1$. \\

The latter can systematically be obtained using \eqref{eq:n-expansion-u} inside
\begin{align}\label{eq:g-pertub}
&\de_x^2 \g_1-2u\de_x\g_1 \left[1+2\sign\g_1(x)\right] +u^2\g_1(x)\left[1+3\sign\g_1(x)+2\g_1^2(x)\right]\nn
&\quad =-u^2 \int_0^1 dy \ \left( \sign +2\g_1(y)\right)\g_2(x,y),
\end{align}
as derived from \eqref{eq:firstorderg} with $\xi(x)\equiv 1$. To first order in $u$, eq.~\eqref{eq:g-pertub} reduces to
\begin{align}\label{eq:g1z}
\de_x^2 \g_1^{\{1\}} = 2\de_x\g_1^{\{0\}}\left[1+2\sign\g_1^{\{0\}}(x)\right] 
\end{align}
with boundary conditions $\g_1^{\{1\}}(0)=\g_1^{\{1\}}(1)=0$. Using \eqref{eq:g0}, this integrates to
\be\label{eq:g1z-sol}
\g_1^{\{1\}}(x)= 2(\bar n_0 -\bar n_N) \left[\frac{1 + 2\sign \bar n_0}{2} x(1-x) + \frac{\sign (\bar n_N - \bar n_0)}{3} x(1-x^2) \right].
\ee
To order ${\cal O}(u^2)$, eq.~\eqref{eq:g-pertub} becomes
\begin{align}\label{eq:g-pertub-2nd}
&\de_x^2 \g_1^{\{2\}}-2\de_x\g_1^{\{1\}}\big(1+2\sign \g_1^{\{0\}}\big) +\g_1^{\{0\}}\left[1+3\sign\g_1^{\{0\}}+2(\g_1^{\{0\}})^2\right]\nn
&=4\sign (\de_x\g_1^{\{0\}})\g_1^{\{1\}} - \int_0^1 dy \ \left( \sign +2\g_1^{\{0\}}(y)\right)\g_2^{\{0\}}(x,y),
\end{align}
with $\g_1^{\{2\}}(0)=\g_1^{\{2\}}(1)=0$ and the unbiased scaling function $\g_2^{\{0\}}(x,y)$ given in \eqref{eq:g2-u=0}. Equation~\eqref{eq:g-pertub-2nd} is recast as
\be
\de_x^2 \g_1^{\{2\}} + C_0 + C_1 x + C_2 x^2 + C_3 x^3=0
\ee
with solution
\be
\g_1^{\{2\}}= \left( \frac{C_0}{2} + \frac{C_1}{6} + \frac{C_2}{12} + \frac{C_3}{20} \right) x- \frac{C_0}{2} x^{2} - \frac{C_1}{6} x^{3} - \frac{C_2}{12} x^{4} - \frac{C_3}{20} x^{5}
\ee
and coefficients
\begin{subequations}
\begin{align}
C_0 &= 
-\tfrac{10}{3} \bar n_0^3 
+ \tfrac{8}{3} \bar n_0^2  \bar n_N 
+ \tfrac{8}{3} \bar n_0 \bar n_N^2 
- \bar n_0 + 2 \bar n_N \nn
&+ \sign\Big(-\tfrac{11}{3} \bar n_0^2 
+ \tfrac{16}{3} \bar n_0 \bar n_N 
+ \tfrac{4}{3} \bar n_N^2\Big); \\[6pt]
C_1 &= 
\tfrac{64}{3} \bar n_0^3 
- 27 \bar n_0^2 \bar n_N 
+ 3 \bar n_0 
+ \tfrac{17}{3} \bar n_N^3 
- 3 \bar n_N \nn
&+ \sign\Big(\tfrac{37}{2} \bar n_0^2 
- 27 \bar n_0 \bar n_N 
+ \tfrac{17}{2} \bar n_N^2\Big); \\[6pt]
C_2 &= 
-27 \bar n_0^3 
+ 54 \bar n_0^2\bar  n_N 
- 27 \bar n_0 \bar n_N^2 \nn
&+ \sign\Big(-\tfrac{27}{2} \bar n_0^2 
+ 27 \bar n_0 \bar n_N 
- \tfrac{27}{2} \bar n_N^2\Big); \\[6pt]
C_3 &= 
9 \bar n_0^3 
- 27 \bar n_0^2 \bar n_N 
+ 27 \bar n_0 \bar n_N^2 
- 9 \bar n_N^3.
\end{align}
\end{subequations}
The third- and fourth-order current cumulants follow from \eqref{eq:F-to-expand} and reads as
\begin{align}
&\frac{\cum_3}{6}=\frac{1}{N}\int_0^1 dx \ \g_1^{\{1\}}(x)\left(1+2\sign \g_1^{\{0\}}(x)\right);\\[4pt]
&\frac{\cum_4}{24}=\frac{1}{N}\int_0^1 dx \ \left[\g_1^{\{2\}}(x)\left(1+2\sign \g_1^{\{0\}}(x)\right) +\sign \left(\g_1^{\{1\}}(x)\right)^2\right].
\end{align}
Explicitly, these are
\be\label{eq:3cum}
\cum_3= \frac{(\bar n_0-\bar n_N)}{N} \left[1+2\sign\,(\bar n_0+\bar n_N)+\frac{4}{15}\left(4 \bar n_0^2+7 \bar n_0 \bar n_N+4 \bar n_N^2\right)\right]
\ee
and
\begin{align}\label{eq:4cum}
\cum_4 &= \frac{1}{N}\Big[\bar n_0+\bar n_N
+ \frac{2\sign}{3}\big(7\bar n_0^2+\bar n_0 \bar n_N+7\bar n_N^2\big) \nn
&\quad + \frac{32 \bar n_0^3+8 \bar n_0^2 \bar n_N+8 \bar n_0 \bar n_N^2+32 \bar n_N^3}{5} \nn
&\qquad + \sign\,\frac{96 \bar n_0^4+64 \bar n_0^3 \bar n_N-40 \bar n_0^2 \bar n_N^2+64 \bar n_0 \bar n_N^3+96 \bar n_N^4}{35}\Big].
\end{align}
Higher-order cumulants can be obtained analogously, although they will involve biased corrections to higher-order $\g_n$ functions. For instance, evaluating $\cum_5$ requires the first bias correction to $\mathfrak{g}_2(x,y)$ given in~\eqref{eq:g2-u1}.\\

Note that the expressions for the first four cumulants obtained here coincide with those of the SSEP for $\sign=-1$~\cite{Derrida2004}. One may also verify that they follow directly from the exact result \eqref{eq:F-exact} through
\be
\cum_q = \frac{1}{N}\,\frac{\de^q}{\de u^q} \E[F_{{\rm qu},w}(u)]\Big\vert_{u=0}.
\ee
\section{Direct computation of {${\rm var}\langle Q(t)\rangle/t$}}\label{app:direct-calc-var}

In this Appendix, we present an alternative derivation of
{${\rm var}\big(\langle Q(t)\rangle\big)/t$} in eq.~\eqref{eq:var-q-ucf}, which does not
rely on the biased Hamiltonian formalism of Sec.~\ref{sec:2time-meas}. This provides an independent
consistency check of the theory developed in the main text. We start
from the unbiased evolution equation for the local density. A direct
computation yields
\begin{equation}
d\hat{n}_{j}=\Delta\hat{n}_{j}\,dt-d\hat{\nu}_{j}+d\hat{\nu}_{j-1},
\end{equation}
where $\Delta$ denotes the discrete Laplacian,
$\Delta f_{j}:=f_{j-1}-2f_{j}+f_{j+1}$, and we have defined
\be
d\hat{\nu}_{j}(t):=-i\left(\hat{\ell}_{j}dw_{j}(t)-\hat{\ell}_{j}^{\dagger}
d\overline{w}_{j}(t)\right).
\ee

Imposing the local conservation law
\begin{equation}
d\hat{n}_{j}=:d\hat{Q}_{j-1}-d\hat{Q}_{j},
\end{equation}
where $\hat{Q}_{j}$ denotes the charge transferred across the link
between sites $j$ and $j+1$, one is naturally led to an
{\it operatorial} definition of the differential of $\hat{Q}_{j}$,
namely
\begin{equation}
d\hat{Q}_{j}=\left(\hat{n}_{j}-\hat{n}_{j+1}\right)dt+d\hat{\nu}_{j}.
\end{equation}
{
Using It\=o calculus,
\be
d(\hat{Q}_{j})^2= 2\hat Q_j \big[\left(\hat{n}_{j}-\hat{n}_{j+1}\right)dt + d\hat \nu_j\big] + (d\hat \nu_j)^2
\ee
yields the stochastic object
\begin{align}
O^{\rm qu}_j:=&\frac{\E\left[d\left(\langle\hat{Q}_{j}^2\rangle_w- \langle \hat Q_j\rangle_w^2\right)\right]}{dt}\nn
=&\E[\langle \hat n_j + \hat n_{j+1}+2\sign \hat n_{j}\hat n_{j+1}\rangle_w]\nn
&  +2\E[\langle\hat Q_j \big(\hat{n}_{j}-\hat{n}_{j+1}\big)\rangle_w-\langle\hat Q_j\rangle_w\langle \hat{n}_{j}-\hat{n}_{j+1}\rangle_w]\nn
&- \frac{\E[\langle d\hat \nu_j\rangle_w^2]}{dt}.
\end{align}
Here $\langle \bullet \rangle_w:=\tr(\hat\rho_{t,w} \bullet)$ denotes the standard (unbiased) quantum expectation. On the other hand, one can also compute the ``classical'' quantity (cf. also Ref.~\cite{Derrida2004})
\begin{align}
O^{\rm cl}_j:&=\frac{d}{dt}\left(\langle \hat Q_j^2 \rangle_{\rm cl}- \langle \hat Q_j\rangle_{\rm cl}^2\right)\equiv\frac{d}{dt}\left(\E[\langle \hat Q_j^2 \rangle_w]- \E[\langle \hat Q_j\rangle_w]^2\right) \nn
& =2 \left(\langle\hat Q_j \left(\hat{n}_{j}-\hat{n}_{j+1}\right)\rangle_{\rm cl} -\langle\hat Q_j \rangle_{\rm cl}\langle\hat{n}_{j}-\hat{n}_{j+1}\rangle_{\rm cl}\right) \nn
&\quad +\langle \hat n_j + \hat n_{j+1}+2\sign \hat n_{j}\hat n_{j+1}\rangle_{\rm cl},
\end{align}
since $\langle \bullet\rangle_{\rm cl}\equiv \E[\langle\bullet\rangle_w]:=\tr(\bar\rho_t \bullet)$ denotes the unbiased classical expectation. The difference $\delta O_j:= O^{\rm qu}_j-O^{\rm cl}_j$ between the two is 
\begin{align}
\delta O_j=&-2\left(\E[\langle \hat Q\rangle_w \langle \hat n_j-\hat n_{j+1}\rangle_w]- \langle \hat Q\rangle_{\rm cl} \langle \hat n_j-\hat n_{j+1}\rangle_{\rm cl} \right)\nn
&-\frac{\E[\langle d\hat \nu_j\rangle_w^2]}{dt}=-\frac{\E[\langle d\hat \nu_j\rangle_w^2]}{dt} +{\cal O}(N^{-2}),
\end{align}
using the axioms \hyperlink{axioms}{({\it iii-iv})} for the factorization of $\E[\ ]$.\\

In the steady state, the current variance is expected to be independent of the link $j$. Therefore, summing over the links, 
\begin{align}
\frac{1}{N}\sum_{j=0}^{N-1}\delta O_j\simeq& -\frac{1}{N}\sum_{j=0}^{N-1}\frac{\E[\langle d\hat \nu_j\rangle_w^2]}{dt}\nn
=& - \frac{2}{N}\sum_{j=0}^{N-1}\E[G^{(w)}_{j,j+1}G^{(w)}_{j+1,j}](0),
\end{align}
with $G^{(w)}_{ji}(u=0)=\langle \hat c^\dagger_j \hat c_i\rangle_w$, one finds
\be
{-N\frac{{\rm var}\left(\left\langle Q(t)\right\rangle\right)}{t}}=\frac{1}{N}\sum_{j=0}^{N-1} \delta O_j \overset{N\to\infty}{\simeq} -\frac{(\bar n_0-\bar n_N)^2}{3N},
\ee
which coincides with the result obtained in eq.~\eqref{eq:res-ucf}.
}

\begin{widetext}

\section{Closure of the equations of motion at any order}\label{app:gn-general}
In this Appendix we show that, upon imposing the gauge condition in Eq.~\eqref{eq:gaugecondition}, the hierarchy of equations closes on $\g_n$ for any $n\geq 1$. We begin by recalling the equation of motion for the two-point function $G^{(w)}_{ij}$. Neglecting boundary terms and suppressing the superscript 
$w$ for clarity, one has
\begin{equation}
dG_{ij}=d{\cal H}_{w,u}\left[G_{i,j}\right] + dt\ {\cal L}_u\left[G_{i,j}\right],
\end{equation}
with
\begin{equation}
 \begin{aligned}{\cal L}_u\left[G_{ij}\right]= & \sum_{k}4\sinh^{2}\left(\frac{u\lambda_{k}}{2}\right)\left(G_{i,k+1}G_{k+1,j}G_{kk}+G_{i,k}G_{k,j}G_{k+1,k+1}\right)+\sign\sum_{k}\left[\left(e^{u\lambda_{k}}-1\right)G_{i,k}G_{k,j}+\left(e^{-u\lambda_{k}}-1\right)G_{i,k+1}G_{k+1,j}\right]\\
 & +\sign\left(e^{u\lambda_{i-1}}-1\right)G_{i-1,i-1}G_{ij}+\sign\left(e^{u\lambda_{j-1}}-1\right)G_{j-1,j-1}G_{ij}+\sign\left(e^{-u\lambda_{i}}-1\right)G_{i+1,i+1}G_{ij}+\sign\left(e^{-u\lambda_{j}}-1\right)G_{j+1,j+1}G_{ij}\\
 & +\left[\left(e^{u\lambda_{i-1}}-1\right)G_{i-1,i-1}+\left(e^{-u\lambda_{i}}-1\right)G_{i+1,i+1}\right]\delta_{ij}-2G_{ij}+\left(G_{i-1,i-1}+G_{i+1,i+1}\right)\delta_{ij}
\end{aligned}
\label{eq:DetG}
\end{equation}
and
\begin{align}
d{\cal H}_{w,u}\left[G_{ij}\right]= & \sign\sum_{k}2i\sinh\left(\frac{u\lambda_{k}}{2}\right)\left(G_{ik}G_{k+1,j}\ d\overline{w}_{k}(t)-G_{i,k+1}G_{kj}\ dw_{k}(t)\right)\label{eq:stochG} \\
 & +ie^{-u\lambda_{j}/2}G_{i,j+1} \ dw_{j}(t)-ie^{u\lambda_{i-1}/2}G_{i-1,j}\ dw_{i-1}(t)+ie^{u\lambda_{j-1}/2}G_{i,j-1}\ d\overline{w}_{j-1}(t)-ie^{-u\lambda_{i}/2}G_{i+1,j}d\overline{w}_{i}(t).
\end{align}
We now consider the $n$-loop of two-point matrices,
\begin{equation}
G^{[n]} := G_{i_{1} i_{2}}\, G_{i_{2} i_{3}} \cdots G_{i_{n} i_{1}},
\end{equation}
where each index $i_j \in [0,N]$. The equation of motion associated with this quantity reads
\begin{align}
dG^{[n]}= & \sum_{j}G_{i_{1},i_{2}}\cdots dG_{i_{j}i_{j+1}}\cdots G_{i_{n-1},i_{n}}+\frac{1}{2}\sum_{j\neq k}G_{i_{1},i_{2}}\cdots dG_{i_{j}i_{j+1}}\cdots dG_{i_{k},i_{k+1}}\cdots G_{i_{n}i_{1}}.
\label{eq:dGn}
\end{align}
By inspecting Eqs.~\eqref{eq:DetG} and \eqref{eq:stochG}, one sees that the
only way to generate a loop of order $n+1$ is through the {insertion of terms $\sum_k(\dots)$ from} the previous expressions. Using It\=o's rule, the
contributions to the first and second terms of Eq.~\eqref{eq:dGn} are,
respectively,
\begin{align}
\sum_{j}G_{i_{1},i_{2}}\cdots\Big\{ & \sum_{k}4\sinh^{2}\left(\frac{u\lambda_{k}}{2}\right)\left(G_{i_{j},k+1}G_{k+1,i_{j+1}}G_{k}^{[1]}+G_{i_{j},k}G_{k,i_{j+1}}G_{k+1}^{[1]}\right)\label{eq:term_1}\\
 & +\sign\sum_{k}\left[\left(e^{u\lambda_{k}}-1\right)G_{i_{j},k}G_{k,i_{j+1}}+\left(e^{-u\lambda_{k}}-1\right)G_{i_{j},k+1}G_{k+1,i_{j+1}}\right]\Big\}\cdots G_{i_{n}i_{1}}\nonumber 
\end{align}
and
\begin{align}
 & \frac{1}{2}\sum_{j\neq k}G_{i_{1},i_{2}}\cdots\sum_{m}2i\sinh\left(\frac{u\lambda_{m}}{2}\right)\left(G_{i_{j},m+1}G_{m,i_{j+1}}dw_{m}(t)-G_{i_{j},m}G_{m+1,i_{j+1}}d\overline{w}_{m}(t)\right)\cdots\\
 &\qquad \sum_{m'}2i\sinh\left(\frac{u\lambda_{m'}}{2}\right)\left(G_{i_{k},m'+1}G_{m',i_{k+1}}dw_{m'}(t)-G_{i_{k},m'}G_{m'+1,i_{k+1}}d\overline{w}_{m'}(t)\right)\cdots G_{i_{n-1},i_{n}}\nonumber \\
= &  \frac{1}{2} dt\sum_{j\neq k}G_{i_{1},i_{2}}\cdots\sum_{m}4\sinh^{2}\left(\frac{u\lambda_{m}}{2}\right)\left(G_{i_{j},m+1}G_{m+1,i_{k+1}}G_{m,i_{j+1}}G_{i_{k},m}+G_{i_{j},m}G_{m,i_{k+1}}G_{i_{k},m+1}G_{m+1,i_{j+1}}\right)\cdots G_{i_{n}i_{1}}\nonumber \\
= & dt\sum_{j\neq k}G_{i_{1},i_{2}}\cdots\sum_{m}4\sinh^{2}\left(\frac{u\lambda_{m}}{2}\right)\left(G_{i_{j},m+1}G_{m+1,i_{k+1}}G_{m,i_{j+1}}G_{i_{k},m}\right)\cdots G_{i_{n}i_{1}}.\nonumber
\end{align}
We see that, because of It\=o's rules, the loop is cut into two smaller
parts, so this term does not generate higher-order loops in the
equations of motion. Hence, the only surviving contribution is given by
Eq.~\eqref{eq:term_1}, which in the continuum limit reads
\begin{equation}
\Lambda\!\left[\mathfrak{g}_{n+1}\right]\!\left(x_{1},\ldots,x_{n}\right).
\end{equation}
Imposing the gauge condition in Eq.~\eqref{eq:gaugecondition} therefore ensures the
closure of the equations of motion on loops of order $n$.

\end{widetext}

\section{Analytic solution of the equations for $\g_1$}\label{app:sol-gauge-eqs}
In this Appendix, we provide details on the exact solution of the set of equations~\eqref{eq:cond1} for the scaling function $\g_1(x)$. Starting from the first equation in \eqref{eq:cond1}, that is
\be\label{eq:app-start}\begin{split}
&\xi(x)\,\partial_x^2 \g_1(x) + 2\,\partial_x \g_1(x)\,\partial_x \xi(x) \\ 
&\quad  -2u\, \g_1(x)\big(1+\sign \g_1(x)\big)\xi(x)\,\partial_x \xi(x) = 0.\end{split}\ee
Following Ref.~\cite{costa2025}, we introduce the auxiliary function $h(x):=\xi(x)\g_1(x)$. In terms of $h(x)$, the previous equation
reduces, after straightforward algebra, to
\be\label{eq:app-SA6}
\de_x^2 h(x) + 2u \sign\, h(x)\,\partial_x h(x) = 0.
\ee
One can check that that eq.~\eqref{eq:app-SA6} is solved by
\be\label{eq:sol-h}
h(x) = \frac{\sign\, A}{\sqrt{u}} \tanh\big(A[x\sqrt{u}+b]\big),
\ee
with free parameters $A$ and $b$. Substituting this expression into the equation \eqref{eq:cond1} for $\xi$ 
gives
\be\begin{split}
\de_x \xi(x) &= -u\,\xi^2(x) -2u\sign \xi(x) h(x)\\
&=-u\,\xi^2(x) -2A\sqrt{u} \xi(x) \tanh\!\big(A[x\sqrt{u}+b]\big).
\end{split}\ee
Integrating this equation,
\be\label{eq:sol-xi}
\xi(x) =\frac{A}{\sqrt{u}}\frac{\text{sech}^2[A(x\sqrt{u}+b)]}{\tanh[A(x\sqrt{u}+b)]+c}, 
\ee
where $c$ is a third integration constant. The three parameters $A$, $b$ and $c$ are fixed by the boundary conditions
\be\label{eq:app-conds}
\begin{cases}
h(0) = \bar n_0 \, \xi(0); \\[6pt]
h(1) = \bar n_N \, \xi(1); \\[6pt]
\int_0^1 dx\ \xi(x) = 1,
\end{cases}\ee
which are, explicitly,
\be\label{eq:conditions}
\begin{cases}\displaystyle
\sign \tanh[Ab]=\bar n_0\frac{\text{sech}^2[Ab]}{\tanh[Ab]+c};\\[12pt]\displaystyle
\sign\tanh[A(\sqrt{u}+b)]=\bar n_N\frac{\text{sech}^2[A(\sqrt{u}+b)]}{\tanh[A(\sqrt{u}+b)]+c};\\[12pt]\displaystyle
\frac{c + \tanh[A(\sqrt{u}+b)]}{c + \tanh[Ab]}=e^u.
\end{cases}
\ee
After some algebra one finds the relations
\begin{align}
\sign\, \bar n_0 (e^u-1) &= \frac{\sinh[Ab] \sinh[A\sqrt{u}]}{\cosh[A(\sqrt{u}+b)]}, \\[4pt]
\sign\, \bar n_N (e^{-u}-1) &= - \frac{\sinh[A\sqrt{u}] \sinh[A(\sqrt{u}+b)]}{\cosh[Ab]} .
\end{align}
A trigonometric identity then gives
\be
\omega(u):=\Big(1-\sign\,\bar n_0(e^u-1)\Big)\Big(1-\sign\,\bar n_N(e^{-u}-1)\Big) 
   = \cosh^2[A\sqrt{u}] .
\ee
Hence
\be
A = \frac{1}{\sqrt{u}}\,\text{cosh}^{-1}\sqrt{\omega(u)}\, .
\ee
By noticing that the cumulant-generating function \eqref{eq:F-general},
\begin{align}\label{eq:app-F}
\E[F_{{\rm qu},w}(u)] &= -u \int_0^1 dx \,\Big( \de_x h(x)+\sign u\, h^2(x)\Big) = -\sign\, u A^2 \nonumber
\end{align}
one immediately finds
\be\label{eq:app-finalF}
\E[F_{{\rm qu},w}(u)] =
-\sign \left( \cosh^{-1} \sqrt{\omega(u)} \right)^2, \quad (\omega(u) \ge 1).
\ee
This expression is well defined for $\omega(u) \ge 1$. The case $\omega(u)\in[0,1]$ follows by analytic continuation and reads
\be
\E[F_{{\rm qu},w}(u)] = \sign \left( \cos^{-1} \sqrt{\omega(u)} \right)^2, \quad (0 \leq \omega(u) \leq 1).
\ee
The critical value $\omega(u) = 1$ is reached when
\be
u_1 = 0, \quad \text{or} \quad
u_2 = \log\left(\frac{\bar n_N\,(1 + \sign\, \bar n_0)}{\bar n_0\,(1 + \sign\, \bar n_N)}\right).
\ee
The nontrivial solution $u_2$ is associated with the Gallavotti-Cohen symmetry, $\E[F_{{\rm qu},w}(u)] = \E[F_{{\rm qu},w}(u_2 - u)]$, as discussed in Sec.~\ref{sec:classical}.\\

For the fermionic case ($\sign = -1$), one has $\omega(u) \ge 0$ for all $u$. On the other hand, for the bosonic case ($\sign = +1$), one can find $\omega(u) < 0$, leading to an unphysical complex-valued cumulant-generating function. This restricts the allowed bias amplitude to $u\in (u_-,u_+)$ with
\be
u_- =- \log\left(1 + \frac{1}{\bar n_N}\right), \quad
u_+ = \log\left(1 + \frac{1}{\bar n_0}\right).
\ee
Outside this interval, the bosonic system undergoes boundary condensation rather than sustaining a large current, see discussion of Sec.~\ref{sec:classical}.\\
\begin{figure}[t!]
\vspace{.1cm}
\centering
\includegraphics[width=.9\columnwidth]{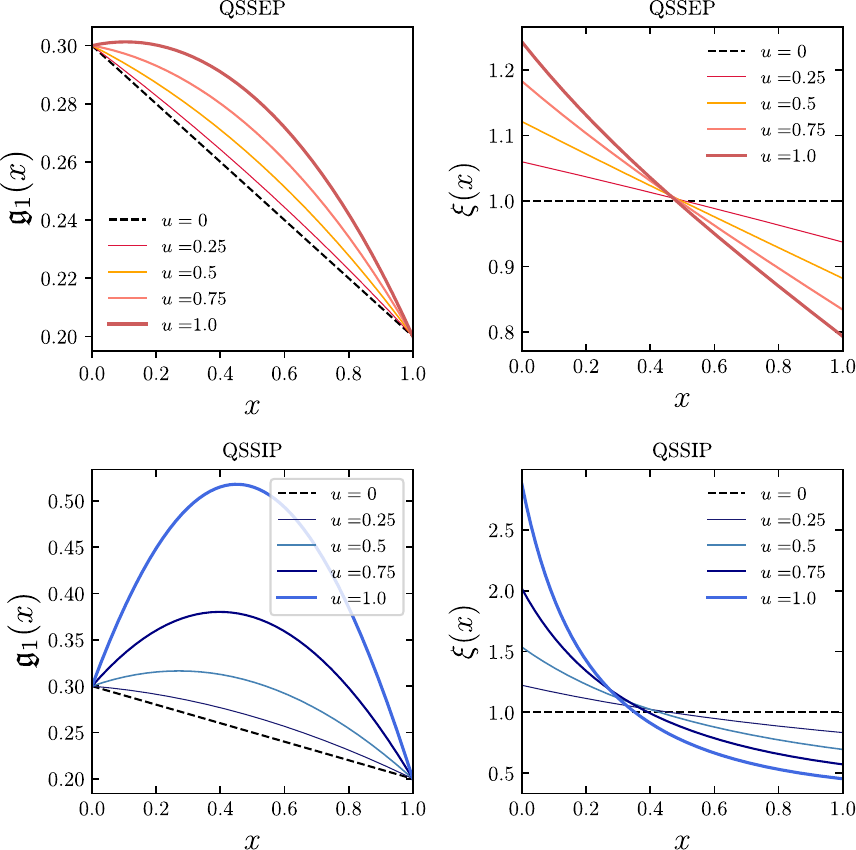}
\caption{Exact solution of the equations in \eqref{eq:cond1} for the functions $\g_1(x)$ (left panels) and $\xi(x)$ (right panels). The plots are made for the fermionic (QSSEP, $\sign=-1$~--~top panels) and bosonic (QSSIP, $\sign=1$,~--~bottom panels) cases, with reservoirs densities fixed to $\bar n_0=0.3$ and $\bar n_N=0.2$.}\label{fig:sol-g-xi}
\vspace{.3cm}
\includegraphics[width=\columnwidth]{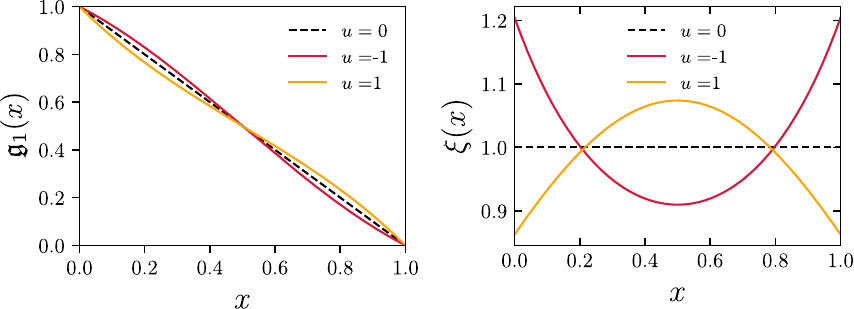}
\caption{Same as Fig.~\ref{fig:sol-g-xi} for the fermionic case (QSSEP, $\sign=-1$) but with reservoirs densities fixed to $\bar n_0=1$ and $\bar n_N=0$. }\label{fig:sol-g-xi-2}
\end{figure}

Eqs.~\eqref{eq:sol-h} and \eqref{eq:sol-xi}, together with the system \eqref{eq:conditions}, fully determine the functions $\g_1$ and $\xi$. Since solving \eqref{eq:conditions} analytically to extract the parameters $b$ and $c$ is rather involved, we determine them numerically. The resulting profiles are shown in Fig.~\ref{fig:sol-g-xi} for the same setup as in Fig.~\ref{fig:F}. We note that these functions may exhibit very different shapes, depending on the particle statistics and on the values of the boundary densities $\bar n_0$, $\bar n_N$. As an illustration, Fig.~\ref{fig:sol-g-xi-2} shows the profiles for the fermionic case with $\bar n_0=1$ and $\bar n_N=0$: since the fermionic density is restricted to the interval $[0,1]$ and the integral of $\xi$ is one, the biased solutions for $\g_1$ and $\xi$ are forced to bend around their unbiased counterparts.

\section{Numerical implementation}\label{app:numerics}
In this Appendix, we provide further details on the numerical implementation of the microscopic dynamics \eqref{eq:dyn-bias}, which was used to generate the data in Figures~\ref{fig:F} and \ref{fig:variance}. The starting point is the biased stochastic update of the biased two-point function $G^{(w)}_{ij}(u; t)=\tr\big(\hat\rho_{t,w}(u)\ \hat c^\dagger_i \hat c_j\big)/Z_{t,w}(u)$, which can be written as
{\begin{align}\label{eq:discr-dyn}
G^{(w)}(u;t+dt)=\tilde{G}^{(w)}(u;t) + dt\  {\cal L}_\text{bdry}[G^{(w)}(u;t)],
\end{align}
with the stochastic bulk update
\begin{align}\label{eq:stoc-update}
&\tilde{G}^{(w)}(u;t)\nonumber\\
&\quad :=e^{idh_{w,-u}} \,\frac{G^{(w)}(u;t)}{\big[1+\sign(1- e^{-idh_{w,u}}e^{idh_{w,-u}})G^{(w)}(u;t)\big]} \, e^{-idh_{w,u}}.
\end{align}
The nonlinear factor in eq.~\eqref{eq:stoc-update} is obtained as follows. Denoting by $e^K$ the Gaussian one-particle kernel introduced in Appendix~\ref{app:gaussian-prop}, one has
\be
G[K]=e^K(1-\sign e^K)^{-1}.
\ee
Under the stochastic bulk update, this kernel transforms as
\be
e^{\tilde K}=e^{i dh_{w,-u}}\,e^K\,e^{-i dh_{w,u}},
\ee
so that $\tilde G=G[\tilde K]$. Rewriting $e^K$ in terms of $G$ yields~\eqref{eq:stoc-update}. Retaining such nonlinear factor in eq.~\eqref{eq:stoc-update} at each stochastic bulk step is important for numerical stability at diffusive times, since finite-step deviations can accumulate.}
The matrices $dh_{w,\pm u}$ above are the biased (non-Hermitian) Hamiltonian in the single-particle basis
\be
d\hat{H}_{w,\pm u}:= \sum_{i,j} \hat{c}^\dagger_i \; dh_{w,\pm u}\; \hat{c}_j.
\ee
The latter can be expressed as the $(N+1)\times (N+1)$ stochastic biased matrix
\begin{align}
&dh_{w,\pm u}:=\nn
&\begin{pmatrix}
0 & {\tt k}_{1,\pm u}(t) & 0 & \cdots & 0 \\
\overline{\tt k}_{1,\pm u}(t) & 0
& {\tt k}_{2,\pm u}(t)& \ddots & \vdots \\
0 & \overline{\tt k}_{2,\mp u}(t )& 0
& \ddots & 0 \\
\vdots & \ddots & \ddots & \ddots
& {\tt k}_{N-1,\pm u}(t) \\
0 & \cdots & 0
& \overline{\tt k}_{N-1,\mp u}(t) & 0
\end{pmatrix}
\end{align}
with
\be
{\tt k}_{j,u}(t):=dw_j(t) e^{\frac{u}{2}\lambda_j}, \quad \overline{\tt k}_{j,u}(t):=d\overline{w}_j(t) e^{\frac{u}{2}\lambda_j},
\ee
and  $(\lambda_0,\dots,\lambda_{N-1})$ satisfying \eqref{eq:gauge}. The (unbiased) boundary terms read as
\be
{\cal L}_\text{bdry}[G]_{i,j} = \sum_{p\in\{0,N\}} \Big[\delta_{i,p}\delta_{i,j}\alpha_p - \frac{\delta_{i,p}+\delta_{j,p}}{2}(\beta_p-\sign\alpha_p)G_{i,j} \Big].
\ee

The discretization \eqref{eq:discr-dyn} leads to a straightforward numerical implementation of the biased quantum dynamics~\eqref{eq:dyn-bias}. The continuous stochastic process is approximated by sampling $dw_j(t)$ at each time step as independent complex Gaussian variables with zero mean and variance $dt$ \cite{Bernard2021}. In our numerical calculations, we approximated the continuous dynamics with a two-step update. We first perform the (stochastic) bulk evolution under the bias in eq.~\eqref{eq:stoc-update}, and subsequently include the (deterministic) boundary evolution with a fourth-order Runge-Kutta integrator (RK4)
\be
G^{(w)}(u;t+dt)=\text{RK4}\big(\tilde{G}^{(w)}(u;t), dt\big).
\ee

Each simulation is performed up to times $t\sim{\cal O}(N^2)$. The robustness of our results with respect to the choice of $dt$ is ensured by checking different scalings of the time step $dt=1/N^k$ ($k=1/2,3/4,1$), for which the global error associated with the boundary evolution scales as ${\cal O}(dt^4 N^2) = {\cal O}(N^{2-4k})$. From the evolution of the biased two-point function, the cumulant-generating function $\E[F_{{\rm qu},w}(u)]$ of Figures~\ref{fig:F}-\ref{fig:variance} is obtained using eq.~\eqref{eq:Ftmp2}.
{\section{Self-averaging}\label{app:details-var}
We provide additional details on the self-averaging estimates \eqref{eq:normalized-log-bounds} and \eqref{eq:var-bound}.  We first discuss the linear stochastic problem in eq.~\eqref{eq:coef}. It is convenient to express the stochastic matrix as
\begin{align}
d\Xi_{nm}(t):=&\tr\big(\hat R_n\,d{\cal H}_{w,u}(\hat L_m)\big)\nonumber\\
=&\sum_{j=0}^{N-1}\Big(A^j_{nm}\,dw_j(t)+B^j_{nm}\,d\overline w_j(t)\Big),
\end{align}
where
\begin{align}
A^j_{nm}&:=-i\Big(e^{u\lambda_j/2}\tr[\hat L_m\hat R_n\hat\ell_j]-e^{-u\lambda_j/2}\tr[\hat L_m\hat\ell_j\hat R_n]\Big),
\\[1mm]
B^j_{nm}&:=-i\Big(e^{-u\lambda_j/2}\tr[\hat L_m\hat R_n\hat\ell_j^\dagger]-e^{u\lambda_j/2}\tr[\hat L_m\hat\ell_j^\dagger\hat R_n]\Big),
\end{align}
with the dependence on the bias $u$ suppressed for notational simplicity. We then separate the contribution of the dominant eigenvalue from the rest by defining
\be
a_t:=c^{(0)}_{t,w},
\qquad
v_{n,t}:=e^{tD_n(u)}c^{(n)}_{t,w},
\quad n>0.
\ee
Eq.~\eqref{eq:coef} then becomes
\begin{align}
da_t&=d\Xi_{00}(t)\,a_t +\sum_{m>0}d\Xi_{0m}(t)\,v_{m,t},
\\
dv_{n,t} &= D_n(u)v_{n,t}\,dt +d\Xi_{n0}(t)\,a_t +\sum_{m>0}d\Xi_{nm}(t)\,v_{m,t}.
\end{align}
These equations show that each nonprincipal mode $v_{n,t}$ is damped by its gap $D_n$, but is also driven by the noise through its coupling to the principal mode $a_t$ and to the other modes $v_{m,t}$. The equations therefore provide no hierarchy allowing us to neglect these fluctuating modes, whose collective contribution enters the finite-size quantum correction. We thus estimate their collective contribution to the growth rate through the normalized state $\hat\Delta_t$ in eq.~\eqref{eq:normalized-state}. Its evolution is
\be
d\hat\Delta_t=d{\cal H}_{w,u}(\hat\Delta_t)+dt\big[({\cal L}_u-\Psi_0(u))\hat\Delta_t\big].
\ee
Applying It\=o calculus to $X_t=\log\tr(\hat\Delta_t)$ gives
\be\label{eq:normalized-log-dynamics}
dX_t=dM_t^\Delta+\big(\mu_t^\Delta+\sigma_t^\Delta\big)dt,
\ee
with
\begin{align}
dM_t^\Delta&:=-2i\sum_{j=0}^{N-1}\sinh\!\left(\frac{u\lambda_j}{2}\right) \left[G^{(w)}_{j+1,j}dw_j-G^{(w)}_{j,j+1}d\overline w_j\right],\label{eq:normalized-martingale}\\
\mu_t^\Delta&:=\frac{\sum_{n>0}D_nv_{n,t}\tr(\hat L_n)}{\tr(\hat\Delta_t)},\label{eq:normalized-drift}\\
\sigma_t^\Delta&:=-4\sum_{j=0}^{N-1}\sinh^2\!\left(\frac{u\lambda_j}{2}\right)G^{(w)}_{j,j+1}G^{(w)}_{j+1,j},\label{eq:normalized-ito}
\end{align}
where $G^{(w)}_{ij}=\tr(\hat\Delta_t\hat c_i^\dagger\hat c_j)/\tr(\hat\Delta_t)\equiv \langle  \hat c_i^\dagger\hat c_j\rangle_{w,u}$. The exact calculation of Sec.~\ref{sec:exact-calc} bounds the mean evolution as
\be\label{eq:normalized-mean-bound}
\frac{d}{dt}\E[X_t]=\E[\mu_t^\Delta+\sigma_t^\Delta]={\cal O}(N^{-2}).
\ee
For the variance,
\be
\frac{d}{dt}{\rm var}(X_t)=2\,{\rm cov}\!\left(X_t,\mu_t^\Delta+\sigma_t^\Delta\right)+\frac{\E[(dM_t^\Delta)^2]}{dt},
\ee
where ${\rm cov}(A,B):=\E[AB]-\E[A]\E[B]$. The martingale contribution satisfies
\be
\frac{\E[(dM_t^\Delta)^2]}{dt}=-2\E[\sigma_t^\Delta]={\cal O}(N^{-2}).
\ee
We next bound the fluctuations of $\sigma_t^\Delta$ and $\mu_t^\Delta$. Writing
\be
q_j:=\sinh^2\!\left(\frac{u\lambda_j}{2}\right)={\cal O}(N^{-2}), \quad F_j^{(w)}:=G^{(w)}_{j,j+1}G^{(w)}_{j+1,j},
\ee
the factorization property (\hyperlink{axioms}{\it iii}) gives
\be
{\rm var}(\sigma_t^\Delta)=16\sum_{j,k}q_jq_k\, {\rm cov}(F_j^{(w)},F_k^{(w)})={\cal O}(N^{-4}).
\ee
To estimate the contribution of the nonprincipal modes, we set $v_{0,t}:=a_t$ and define
\be
\alpha_{n,t}:=\frac{v_{n,t}\tr(\hat L_n)}{\tr(\hat\Delta_t)}
\ee
such that $\sum_n\alpha_{n,t}=1$, and
\be
\mu_t^\Delta=\sum_{n>0}D_n\alpha_{n,t}.
\ee
As anticipated in the main text, we assume that the modes relevant on diffusive time scales have gaps $D_n=d_n/N^2$, while the corresponding weighted sum $\sum_{n>0}d_n\alpha_{n,t}$ remains bounded in mean square. It follows that
\be
{\rm var}(\mu_t^\Delta)={\cal O}(N^{-4}).
\ee

Cauchy-Schwarz inequality then gives
\be
\left|{\rm cov}\left(X_t,\mu_t^\Delta+\sigma_t^\Delta\right)\right|\leq\frac{C}{N^2}\sqrt{{\rm var}(X_t)},
\ee
and using $\sqrt{x}\leq 1+x$, we obtain
\be\label{eq:dvar}
\frac{d}{dt}{\rm var}(X_t)\leq \frac{C}{N^2}\big(1+{\rm var}(X_t)\big),
\ee
where $C$ is some constant independent of $N$. For a normalized initial state, $X_0=0$. Integrating eq.~\eqref{eq:normalized-mean-bound} and applying Gronwall's inequality to eq.~\eqref{eq:dvar} up to $t=\tau N^2$ proves eq.~\eqref{eq:normalized-log-bounds}, and hence the self-averaging bound in eq.~\eqref{eq:var-bound}.}
\section{Extremization conditions from variational ansatz}\label{app:details-extr-variational-app}
We provide additional details for the extremization \eqref{eq:var-ansatz-extr} of the functional ${\cal F}(\hat L,\hat R)$ in eq.~\eqref{eq:var-func-def}. Evaluating the functional on Gaussian trial states \eqref{eq:trial-states} leads to the expression
\be\label{eq:func-explicit}
{\cal F}_u(\hat{L},\hat{R}) = \sum_{j=0}^{N-1} \mathscr{F}_{j,j+1}({\rm f},{\rm e})
\ee
with
\begin{align}\label{eq:F-gauss}
\mathscr{F}_{j,k} :=& e^{-u\lambda_j} \Big[(\e^{-1} G\e)_{k,j} G_{j,k}  +(\e^{-1} G)_{k,k}\left(\e(\mathbb{I}+\sign G)\right)_{j,j}\Big]\nn
&+e^{u\lambda_j} \Big[(\e^{-1T} G \e^T)_{j,k} G_{k,j} + (\e^{-1T}G)_{j,j}\left(\e^T(\mathbb{I}+\sign G)\right)_{k,k}\Big]\nn
&-\frac12\Big[(\e^{-1} G \e)_{j,k}(\e^{-1T}G\e^T)_{k,j} +(\e^{-1} G \e)_{k,j}(\e^{-1T}G\e^T)_{j,k} \nn
&+ (\e^{-1} G \e^T)_{j,j}\left(\e^{-1}(\mathbb{I}+\sign G)\e^T\right)_{k,k} + G_{j,j}(1+\sign G_{k,k})\nn
&+ (\e^{-1} G \e^T)_{k,k}\left(\e^{-1}(\mathbb{I}+\sign G)\e^T\right)_{j,j} +G_{k,k}(1+\sign G_{j,j})\Big]\nn
&-G_{j,k}G_{k,j}+ \text{bdry},
\end{align}
with notation $\e^{-1T}\equiv (\e^{-1})^T$. Here, ${\rm e} := \exp(K^{(R)})$, ${\rm f} := \exp(K^{(L)})$,  and the two-point function
\be
G_{k,j}({\rm f},\e) := \frac{\tr\big(\hat{L}\hat{R} \ \hat{c}^\dagger_k \hat{c}_j\big)}{\tr(\hat{L}\hat{R})} = \left( \frac{{\rm f} \, {\rm e}}{1 - \sign\  {\rm f}\, {\rm e}} \right)_{k,j}.
\ee
For conciseness, we omit the explicit form of the boundary terms. We proceed by computing the variations $\delta_{\ \e} {\cal F}_u$ and $\delta_{\ {\rm f}} {\cal F}_u$ of \eqref{eq:func-explicit} using Fréchet calculus. For instance,
\be
\delta_\e \big( \e^{-1} G \e\big)=-\e^{-1} \delta\e\  \e^{-1} G \ \e + \e^{-1} (\delta_\e G) \e + \e^{-1} G \ \delta\e
\ee
and so on for each term in \eqref{eq:func-explicit}. After some algebra, one obtains explicit expressions for the two variations. We do not report them here, as they consist of many terms and lead to rather cumbersome formulas. The extremum condition,
\be\label{eq:extr-general}
\begin{cases}\displaystyle
\frac{\delta {\cal F}_u}{\delta\e}\Big\vert_{{\rm f}_*, \e_*}= 0;\\[.5cm]\displaystyle
\frac{\delta {\cal F}_u}{\delta{\rm f}}\Big\vert_{{\rm f}_*, \e_*} = 0,
\end{cases}
\ee
{specifies $\e_*$ and ${\rm f}_*$ in the ensemble of Gaussian trial states}. By restricting our analysis to the subspace of {\it diagonal} Gaussian states, namely by looking for $({\rm f}_*,\e_*)$ of the form
\be\label{eq:diagonal-saddle}
{\rm f}_*= \e_*=:\text{diag}(e^{\phi})\quad \text{such that}\quad  \phi_N-\phi_0=0,
\ee
the extremum conditions \eqref{eq:extr-general} yield the set of equations
\begin{subequations}\label{eq:extr-cond-last}
\begin{align}\label{eq:extr-equation}
&e^{u\lambda_{k}} n_{k+1}(1+\sign n_k) -e^{u\lambda_{k-1}} n_k(1+\sign n_{k-1}) \nonumber\\
&\quad - e^{-u\lambda_{k}} n_k(1+\sign n_{k+1}) + e^{-u\lambda_{k-1}} n_{k-1}(1+\sign n_{k}) \nonumber\\[1mm]
&= 2 n_k (1+\sign n_k) \Big[2\sign  (n_{k+1} + n_{k-1}) + 2
 -\sign (e^{u\lambda_k} n_{k+1}) \nn &
 \quad - e^{u\lambda_{k-1}} (1 +\sign n_{k-1}) -\sign (e^{-u\lambda_{k-1}} n_{k-1}) - e^{-u\lambda_k} (1+\sign n_{k+1}) 
 \Big]
\end{align}
and
\begin{align}
&\big[e^{u\lambda_{k-1}}-1\big]\left(1+\sign n_{k-1}\right)+ \big[e^{-u\lambda_{k}}-1\big]\left(1+\sign n_{k+1}\right) \nn
&+ \sign \big[e^{u\lambda_{k}}-1\big]n_{k+1} + \sign \big[e^{-u\lambda_{k-1}}-1\big]n_{k-1}=0,
\end{align}
\end{subequations}
where $n_k({\rm f},\e):=G_{kk}({\rm f},\e)$, and we redefined $\lambda_k - (\phi_{k+1}-\phi_k) \equiv \lambda_k$, exploiting the arbitrariness of $\lambda_k$.\\

From eqs.~\eqref{eq:extr-cond-last}, by expanding the lattice functions $n_{j\pm1}$ and $\lambda_{j\pm1}$ at large $N$ in terms of the continuous variable $x=j/N\in[0,1]$, one recovers the set of equations in \eqref{eq:cond1}. Consistently with this, evaluating \eqref{eq:F-var} at the diagonal extremum~\eqref{eq:diagonal-saddle} shows that, to leading order in $1/N$, the cumulant-generating function $F_{{\rm qu},w}(u)$ has the same expression as $F_{\rm cl}(u)$ in \eqref{eq:F-general}. \\

This finding supports the conjectured classical typicality of the quantum process in the large-$N$ limit, i.e.,
\begin{equation}
F_{{\rm qu},w}(u)\,\overset{N\to\infty}{\simeq}\, F_{\rm cl}(u).
\end{equation}

{It should be noted that, although the choice in \eqref{eq:diagonal-saddle} yields a well-defined extremum of the functional~\eqref{eq:var-func-def}, other extrema may exist for nondiagonal choices of $({\rm f}_*,\e_*)$. A further exploration of such possibilities could clarify their possible role in the subleading terms and the relation with the subleading quantum correction $\delta F_{\rm qu}$.}
\bibliographystyle{apsrev4-2}
\bibliography{bibliography_v2}

\end{document}